\documentclass[conference]{IEEEtran}
\IEEEoverridecommandlockouts
\usepackage{cite}
\usepackage{amsmath, amssymb, amsfonts, amsthm, bm, blkarray, comment, mathtools, enumitem, adjustbox}
\usepackage{algorithmic}
\usepackage{graphicx}
\usepackage{textcomp}
\def\BibTeX{{\rm B\kern-.05em{\sc i\kern-.025em b}\kern-.08em
    T\kern-.1667em\lower.7ex\hbox{E}\kern-.125emX}}

\usepackage[colorlinks=true, allcolors=blue]{hyperref}
\usepackage{breakurl}
\usepackage{booktabs}
\usepackage{lipsum}
\usepackage[table]{xcolor}
\usepackage{libertine}
\usepackage{tcolorbox}
\usepackage[subrefformat=parens]{subcaption} 
\usepackage{tikz}
\usepackage{pgfplots}
\usepackage{multirow}

\begin{document}
  
\title{Cryptoeconomics and Tokenomics as Economics: \\A Survey with Opinions
}

\author{
\IEEEauthorblockN{Kensuke ITO}
\IEEEauthorblockA{\textit{Endowed Chair for Blockchain Innovation} \\
\textit{The University of Tokyo}\\
Tokyo, Japan \\
k-ito@g.ecc.u-tokyo.ac.jp}
}

\maketitle

\begin{abstract}
This paper surveys products and studies on {\em cryptoeconomics} and {\em tokenomics} from an economic perspective, as these terms are still (i) ill-defined and (ii) disconnected from economic disciplines. 
We first suggest that they can be novel when integrated; we then conduct a literature review and case study following {\em consensus-building for decentralization} and {\em token value for autonomy}.
Integration requires simultaneous consideration of strategic behavior, spamming, Sybil attacks, free-riding, marginal cost, marginal utility and stabilizers.
This survey is the first systematization of knowledge on cryptoeconomics and tokenomics, aiming to bridge the contexts of economics and blockchain.
\end{abstract}

\section{Introduction}

{\em Bitcoin} \cite{nakamoto2008bitcoin} is the first practical protocol that employed an economic incentive design to implement a peer-to-peer electronic cash system.\footnote{Peer-to-peer electronic cash systems prior to the Bitcoin protocol include, for example, {\em b-money} \cite{dai1998b} and {\em Bit gold} \cite{szabo2005bit}. Comprehensive histories of the pre-Bitcoin era are provided in the literature \cite{brunton2020digital, gladstein_2021_the}.}
Specifically, it has enabled consensus-building on transaction records among an unspecified number of strategic peers\textemdash a long-standing problem for peer-to-peer electronic cash\textemdash with cryptography and subject to the following main rules:

\begin{tcolorbox} [title={\em Main Rules of the Bitcoin Protocol}]
\begin{itemize}
    \item Transaction records are sequentially stored in blocks, and peers maintain an identical chain of blocks through consensus-building ({\em blockchain}).
    \item Peers can create a new block and attach it to any existing block in the chain; however, the success of this task is probabilistic, directly proportional to the relative amount of computing resources expended by the peer ({\em proof-of-work} \cite{dwork1992pricing, jakobsson1999proofs}).  
    \item In the event of a chain fork into multiple paths, the longest chain is accepted as the consensus ({\em Nakamoto consensus}). 
    \item Peers who successfully create a block in the longest chain receive newly minted Bitcoins as rewards ({\em coinbase as contribution rewards}).
\end{itemize}
\end{tcolorbox}

\noindent
In other words, the Bitcoin protocol discourages undesirable actions by making them unprofitable rather than impossible.
The protocol's novelty lies in its use of economic incentives for decentralized autonomous consensus-building.

This novelty has been carried forward to subsequent blockchain-related products, such as {\em Ethereum} \cite{buterin2014ethereum, wood2014ethereum} (a protocol that extended consensus-building to cover everything from transaction records to state transitions, thereby facilitating {\em decentralized applications} [DApps] \cite{johnston2014thegeneraltheoryofdecen}), {\em The DAO} \cite{jentzsch2016decentralized} (one of the first DApps on Ethereum, designed to function as a decentralized autonomous investment-fund in Ether),\footnote{The term {\em decentralized autonomous organization} (DAO) first introduced in the Ethereum Whitepaper as ``a virtual entity that has a certain set of members or shareholders which, perhaps with a 67\% majority, have the right to spend the entity's funds and modify its code \cite{buterin2014next}.''
Buterin also described a concept closely resembling what is now known as a DAO in 2013:  
``But what if, with the power of modern information technology, we can encode the mission statement into code; that is, create an inviolable contract that generates revenue, pays people to perform some function, and finds hardware for itself to run on, all without any need for top-down human direction? \cite{buterin2013bootstrapping}.''
Note that The DAO was hacked in 2016 and and is currently inaccessible \cite{santos2018dao, morrison2020dao}. 
The current characteristics and types of DAOs are detailed in the literature \cite{wang2019decentralized, el2020overview, ding2023survey}.}
and {\em Lightning Network} \cite{poon2015bitcoin} (an additional layer on the Bitcoin protocol aimed at increasing transaction processing speed, i.e., scalability).

\begin{figure}[t]
\centering
 \includegraphics[width=0.85\hsize]{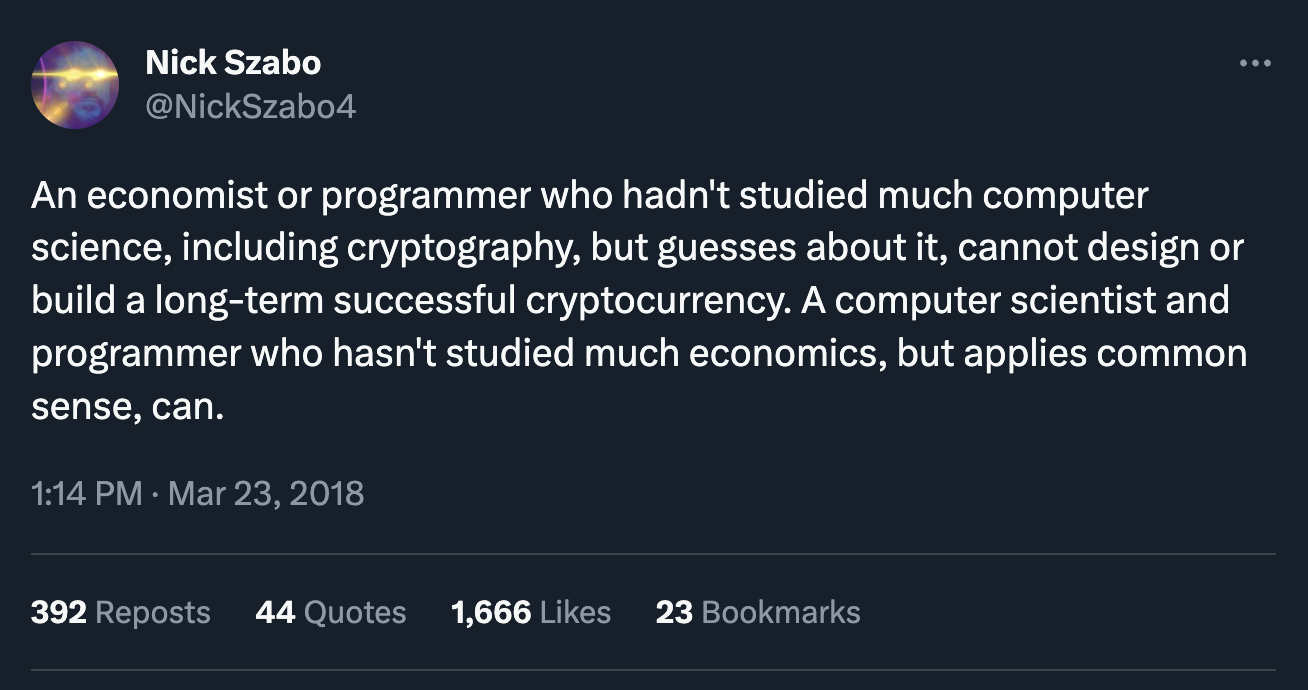}
 \caption{Disconnect between Economics and Blockchain}
 \captionsetup{font=footnotesize, justification=raggedright, singlelinecheck=off}
    \caption*{{\em Source}: X.com (\url{https://x.com/NickSzabo4/status/977035747713675264}, accessed September 4, 2023).}
    \captionsetup{font=normal}
 \label{szabo}
\end{figure} 

{\em Cryptoeconomics} and {\em tokenomics} emerged from the above contexts (c. 2014-2016).
Namely, if taken in purpose rather than definition, these terms extend the coverage of Bitcoin's novelty\textemdash utilizing economic incentives for decentralized autonomous consensus-building\textemdash from on-chain transaction records to more generalized information, encompassing on-chain state transitions (Ethereum), off-chain peer beliefs on appropriate investments (The DAO), and off-chain transaction records (Lightning Network).

Despite their importance, cryptoeconomics and tokenomics face two main challenges at the time of this writing.
First, they are ill-defined. 
Section \ref{history} illustrates how these terms are often used without clear definitions or distinctions, reflecting their emerging and interdisciplinary nature \cite{voshmgir2019foundations}.
This ambiguity makes productive discussions difficult.
Second, even with the name `economics,' they are disconnected from economic disciplines.
As cryptographer Nick Szabo's post (Figure \ref{szabo}) implies, these terms have not sufficiently referred to economics or prior attempts to integrate economics and computer science (see Sections \ref{consensus} and \ref{token} for more details). 
Conversely, economists have shown only limited (or one-sided) interest in cryptoeconomics and tokenomics, despite the potential for significant synergies between economics and blockchain-related products.
This disconnect, a factor in reinventing the wheel, further makes productive discussions difficult.

This paper surveys products and studies behind cryptoeconomics and tokenomics from an economic perspective to address the two challenges.
We address the first challenge by proposing a new definition through a historical review of the terms; the second challenge is examined via a literature review and case studies that explore {\em consensus-building for decentralization} and {\em token value for autonomy}.
To the best of the author's knowledge, this survey represents the first systematization of knowledge (SoK) on cryptoeconomics and tokenomics that aims to bridge the respective contexts of economics and blockchain.\footnote{Strictly speaking, this survey is an extended version of the report by Ito (2018) \cite{ITO_2018} (in Japanese).}

This paper is organized into six sections, including this introduction.
Section \ref{history} covers history and opinions for terminology, 
Section \ref{consensus} reviews prior studies on designing consensus-building for decentralization, and
Section \ref{token} reviews prior studies on designing token value for autonomy, 
Section \ref{case} provides case studies for each protocol and DApp, and
Section \ref{conclusion} presents the conclusion and future research directions.

\section{History and Opinions for Terminology} \label{history}

Cryptoeconomics and tokenomics serve clear purposes, yet their definitions and distinctions remain ambiguous. 
For productive discussion, this section will first chronologically review the history of these two terms and then present the author's opinion on their definitions.

\subsection{Cryptoeconomics} \label{hiscry}

The term cryptoeconomics originated within the Ethereum community and was first publicly used by Vlad Zamfir, a core member of the {\em Ethereum Foundation} \cite{ethereumHomeEthereum}, in a 2015 presentation.
Zamfir defined cryptoeconomics as
``a formal discipline that studies protocols that govern the production, distribution and consumption of goods and services in a decentralized digital economy. Cryptoeconomics is a practical science that focuses on the design and characterization of these protocols \cite{zamfir_2015_what}."

In the same year, Vitalik Buterin, co-founder of Ethereum, described the concept {\em cryptoeconomic} as follows: ``it's decentralized, it uses public key cryptography for authentication, and it uses economic incentives to ensure that it keeps going and doesn't go back in time or incur any other glitch \cite{buterin_2015_visions}."
Buterin further refined this concept in a 2017 presentation, stating that cryptoeconomics involves ``Building systems that have certain desired properties. 
Use cryptography to prove property about messages that happened in the past.
Use economic incentives defined inside the system to encourage desired properties to hold into the future \cite{buterin_2017_introduction}."
Here, he discussed examples such as {\em SchellingCoin} \cite{Buterin_2014} and blockchain forks, which leverage cryptography and game-theoretic coordination for consensus-building. 

\begin{figure}[h]
\centering
\includegraphics[width=1\hsize]{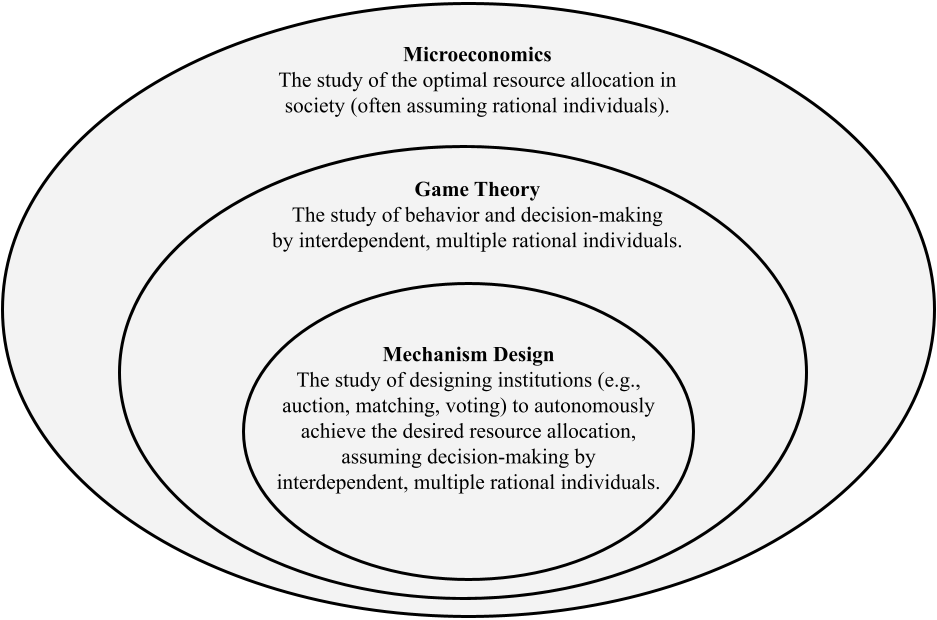}
\caption{Relationship between Microeconomics, Game Theory, and Mechanism Design}
\label{eco}
\end{figure}

Davidson, et al. (2016) \cite{davidson2016economics} was among the first publication in academia to explicitly mention cryptoeconomics, positioning it as ``a branch of mechanism design, which is a branch of microeconomics \cite{davidson2016economics}," based on the definition by Zamfir.
Obasi (2017) \cite{obasi_2017_the} similarly described cryptoeconomics as follows: ``It has more in common with mechanism design\textemdash an area of mathematics and economic theory, sometimes referred to as reverse game theory \cite{obasi_2017_the}." 
Furthermore, Voshmgir and Zargham (2019) \cite{voshmgir2019foundations} referred to cryptoeconomics as ``an emerging field of economic coordination games in cryptographically secured peer-to-peer networks \cite{voshmgir2019foundations}," highlighting its interdisciplinary connections with system theory, political science, and network science.
See Figure \ref{eco} below for the relationship between microeconomics, game theory, and mechanism design that the series of discussions implicitly assume.

Finally, Brekke and Alsindi (2021) \cite{brekke2021cryptoeconomics} synthesized these discussions and provided the following general definition: 
``Cryptoeconomics describes an interdisciplinary, emergent and experimental field that draws on ideas and concepts from economics, game theory and related disciplines in the design of peer-to-peer cryptographic systems. Cryptoeconomic systems try to guarantee certain kinds of information security properties using incentives and/or penalties to regulate the distribution of efforts, goods and services in new digital economies \cite{brekke2021cryptoeconomics}."

Thus, cryptoeconomics focuses on economic incentives as a means of consensus-building and, although interdisciplinary, it would be placed in the context of mechanism design.\footnote{Brekke and Alsindi (2021) \cite{brekke2021cryptoeconomics} argues the interdisciplinary nature of cryptoeconomics as follows: ``Cryptoeconomics is an embryonic field at present and can be taken to include several areas of focus: information security engineering, mechanism design, token engineering and market design \cite{brekke2021cryptoeconomics}."}

\subsection{Tokenomics} \label{histoke}

The term tokenomics has been in use since at least 2012 \cite{au2018tokenomics}; to the author's knowledge, Mougayar (2017) \cite{mougayar_2017_startup} first applied it in blockchain discussions.
Unlike many other studies, he interpreted cryptoeconomics as ``a term that has come to describe the mechanics and specifics of token distribution, according to a given sale and ownership structure \cite{mougayar_2017_startup}," and introduced the term tokenomics as a contrast to emphasize the importance of ``the utility role of the token \cite{mougayar_2017_startup}" to ``deliver a viable business model for the long term \cite{mougayar_2017_startup}."
In other words, tokenomics was originally used to promote attention to tokens' demand side (i.e., utility) rather than their supply side.

In contrast, Ennis, et al. (2018) \cite{weaver_2018_three} mentioned three aspects of tokenomics: ``(1) a means of self-funding within the crypto economy, (2) the deployment of a token within the ecosystem of an ICO project and (3) the set of all economic activity generated through the creation of tokens \cite{weaver_2018_three}."
Here, tokenomics was extended to cover both the demand and supply sides.\footnote{Blemus and Guégan (2020) \cite{blemus2020initial} interpret this definition as ``a self-funding mechanism for projects within the crypto economy \cite{blemus2020initial}" and add the following two to this primary definition:
``would apprehend the notion of token, beyond this strictest definition, along the lines of its function \cite{blemus2020initial}," ``tokenomics which focuses on the economic activity and value generated through the token creation \cite{blemus2020initial}."}

Au and Power (2018) \cite{au2018tokenomics} defined tokenomics even more broadly, suggesting an overlap with cryptoeconomics.
They described tokenomics as ``the concept of the study, design, and implementation of an economic system to incentivize specific behavior in a community, using tokens to create a self-sustaining ad hoc mini-economy. It includes game-theory, mechanism design and monetary economics \cite{au2018tokenomics}." 

Finally, Kampakis (2022) \cite{kampakis2022auditing} offered the broadest definition, characterizing tokenomics as
``the study of how crypto tokens are used within the blockchain ecosystem \cite{kampakis2022auditing},"
which encompasses:
``1) The number of tokens issued and the way they are issued (vesting schedule, airdrops, etc.).
2) The economics of a consensus algorithm; largely referred to as crypto-economics.
3) The general structure of the system: game theoretic and economic incentives \cite{kampakis2022auditing}."

As shown, tokenomics began with a discussion of token value and gradually came to encompass cryptoeconomics as a subcategory.

\subsection{Opinion: They Are Individually Not So Novel} \label{oppre}

Such trends lead to two opinions. 
First, cryptoeconomics and tokenomics have similar precedents in economics.

For cryptoeconomics, game theory was applied to cryptography at least as early as 1993 \cite{fischer1993application}, and the two has been studied together ever since \cite{katz2008bridging}.
Nisan et al. (1999) \cite{nisan1999algorithmic} and subsequent studies \cite{nisan2001algorithmic, nisan2007algorithmic} have developed an {\em algorithmic mechanism design} (AMD) that applies mechanism design to computational issues (e.g., routing and load balancing), considering additional constraints like computational resources.
Furthermore, as a branch of AMD, Feigenbaum et al. (2000) \cite{feigenbaum2000sharing} even proposed the concept of {\em distributed algorithmic mechanism design} (DAMD), which primarily focuses on peer-to-peer systems where agents, computational resources, and networks are all distributed.
A survey paper \cite{feigenbaum2004distributed} in 2004 outlined several open problems in DAMD, including the following:

\vspace{0.4\baselineskip}
``Open Problem 10. Can digital signatures (or, more generally, cryptographic protocol-design techniques) always be used to convert a distributed algorithmic mechanism in which some of the parties must be assumed to be obedient into one with a more realistic strategic model? \cite{feigenbaum2004distributed}"

\vspace{0.4\baselineskip}
``Open Problem 18. Can one design monetary P2P systems that provide better performance than purely barter P2P systems? Can one characterize, in simple models, the possible outcomes achievable with both kinds of P2P economies? \cite{feigenbaum2004distributed}"

\vspace{0.4\baselineskip}
\noindent
Although raised a decade before the implementation of Ethereum, these open problems are very close to those in cryptoeconomics.

For tokenomics, value and price were theorized in 19th-century (neoclassical) economics \cite{marshall1890principles}, which states that a good's market price is determined as the intersection of supply-side value (depends on marginal cost) and demand-side value (depends on marginal utility).\footnote{Marshall (1890) \cite{marshall1890principles} likened the debate over whether the source of value lies in cost or utility to a futile debate over whether the top or bottom blade of scissors does the cutting.
This perspective is relevant in discussing the Bitcoin's value, where the marginal cost corresponds to the computational resource of proof-of-work, and the marginal utility corresponds to its utility as an electronic peer-to-peer cash system and transaction fee.
The market price of Bitcoin is at their intersection (see Section \ref{token} for more details). 
Despite the theory's simplicity, whether Bitcoin's value lies in proof-of-work is still often debated, and to the author's knowledge, no economists has pointed out the futility of the debate. 
This situation reflects the disconnect between blockchain and economics.}
Based on this theory, economic studies have analyzed more specific goods, such as money \cite{handa2008monetary}, securities, and stocks \cite{rubinstein2011history}, and developed dynamic models that consider individual expectations.
These studies share similarities with tokenomics \cite{conley2017blockchain} and have been extended to the design of (rule-based) monetary policies \cite{taylor1999historical} and digital currencies \cite{raskin2016digital}.
These precedents are very close to tokenomics because they are trying to design financial rules and products based on theories of value and price.

\subsection{Opinion: They Can Be Novel When Integrated} \label{opinte}

Second, cryptoeconomics and tokenomics can be novel when integrated.
Despite similar precedents for each, to the author's knowledge, no academic efforts have addressed (cryptoeconomic) consensus-building and (tokenomics) token value together; however,
such integration is essential because the two aspects are interrelated in practice.
Consensus-building cannot be autonomous if the token as a reward does not have value,
and token value does not contribute to decentralization if the product lacks consensus-building.
Accordingly, cryptoeconomics and tokenomics should be integrated and renamed to reflect {\em the design of mechanism and reward}, or {\em token-based mechanism design}. 

\vspace{0.4\baselineskip}
This section organized the history of cryptoeconomics and tokenomics chronologically, suggesting that they can be novel when integrated.
The next two sections will present blockchain-related products and prior studies in more detail, following two categories for integration: {\em designing consensus-building for decentralization} (Section \ref{consensus}) and {\em designing token value for autonomy} (Section \ref{token}).

\section{Designing Consensus-Building for decentralization} \label{consensus}

In the context of blockchain, decentralization is strongly associated with consensus-building;\footnote{Here, the usage of the term decentralization follows Hoffman et al. (2020) \cite{hoffman2020toward} and Zhang et al. (2022) \cite{zhang2022sok}.}
instead of delegating centralized authorities, we require some mechanism for eliciting and aggregating information from peers and making some output as a consensus (e.g., the aforementioned {\em main rules of the Bitcoin protocol}).
This section presents what obstacles blockchain-related products and prior studies have faced and addressed in designing such consensus-building for decentralization.

\subsection{How to Address Strategic Behavior} \label{strategic}
First, consensus-building must prevent the strategic behavior of peers, which intuitively denotes the action of deliberately misreporting accurate information.\footnote{In computer science, strategic behavior encompasses a broader concept known as the {\em Byzantine Generals Problem} \cite{lamport1982byzantine}, which includes considerations of unintentional malfunctions of peers, such as communication failures. The Bitcoin protocol offers a quasi-solution to this problem, given that Nakamoto consensus lacks finality (\ref{norm}).}
For example, peers in the Bitcoin protocol might create a block containing conflicting transactions (e.g., double-spending);
peers in The DAO might misreport their beliefs about appropriate investments to gather more Ether to their proposals.
Such behavior would be more likely to occur given the possibility of collusion among multiple peers.

The Bitcoin protocol addresses strategic behavior by adopting the aforementioned main rules.
Particularly, the combination of proof-of-work and Nakamoto consensus turns consensus-building (on transaction records) into a form of majority voting, where voting power is proportional to each peer's computational resources.
Furthermore, many subsequent blockchain-related products have adopted variations of the following {\em token-staking} rules for consensus-building.

\begin{tcolorbox} [title={\em Token Staking (a simple example of binary choice)}]
    \begin{itemize}
        \item Peers can stake any amount of their tokens to either {\em accept} or {\em reject} a proposal, 
        \item Consensus is the choice of which collects more tokens after a certain period, 
        \item All staked tokens are redistributed among peers who staked them on the consensus side. 
    \end{itemize}
\end{tcolorbox}

\noindent
Namely, token-staking is assumed to prevent strategic behavior by i) majority voting with tokens (mostly at some cost to obtain) and ii) the penalty of losing staked tokens.
The above is a simple example, and various forms of token-staking exist for consensus-building.
The DAO adopted the token-staking with multiple choices.
{\em Nouns DAO} \cite{nounsNouns}, another DAO detailed in \ref{tokendemand} and \ref{casenouns}, adopts token-voting, where staked-tokens are neither redistributed nor burned (i.e., trying to prevent strategic behavior without penalty).
Ethereum adopts {\em proof-of-stake} \cite{king2012ppcoin} and {\em Gasper} \cite{buterin2020combining}, where staked tokens are not redistributed but are burned when a peer misreports (i.e., staked tokens are like a deposit for consensus-building).\footnote{Proof-of-stake is responsible for selecting validators, while Gasper aggregates their opinions for consensus-building. Specifically, proof-of-stake probabilistically selects both the validators who propose new blocks and those who vote on proposed blocks, based on the amount of Ether they have staked. Gasper, on the other hand, combines two mechanisms ({\em LMD GHOST} and {\em Casper}) to achieve consensus on the legitimate chain in the event of a blockchain fork, using votes and staked Ether. 
Ethereum transitioned from proof-of-work/Nakamoto consensus to proof-of-stake/Gasper in its September 2022 update, aiming to address norms such as finality, energy efficiency, and scalability (which are discussed in \ref{norm}). For detailed information on Ethereum's new consensus-building, see Pavloff et al. (2023) \cite{pavloff2023ethereum}.}

Prior studies in economics, especially in game theory, have formalized strategic behavior and its solution concepts.
For example, {\em strategy-proofness} ({\em truthfulness}) is a solution concept where players in a {\em mechanism} \cite{parkes2004learnable} cannot gain more utility by deviating from truth-telling.\footnote{Note that strategy-proofness ensures truthtelling is a weakly dominant strategy, meaning that it allows for cases where truth-telling and other strategy yield the same amount of utility. More practical solution concept, such as {\em strongly truthfulness} \cite{dasgupta2013crowdsourced}, also exist. See Tardos and Vazirani (2007) \cite{tardos2007basic} for example about the detail and differences of other solution concepts.} 
Strategy-proofness has been applied to systems including voting, and the mechanism satisfying it was later generalized as the {\em VCG mechanism} \cite{vickrey1961counterspeculation, clarke1971multipart, groves1973incentives}.\footnote{Mechanism design is often referred to as inverse game-theory because it derives institutions from solution concepts, not solution concepts from institutions.}
It would be natural to apply the game-theoretic approach to blockchain-related products.
In particular, a number of studies have modeled the Bitcoin protocol as a game and analyzed peers' activities using solution concepts like strategy-proofness \cite{liu2019survey, halaburda2022microeconomics, warren2023bitcoin}.
Token-staking has also been modeled in game theory \cite{saleh2021blockchain}. 
These studies, especially the axiomatic ones \cite{chen2019axiomatic, leshno2020bitcoin}, tend to evaluate the combination of proof-of-work and Nakamoto consensus compared to others based on token-staking.

The {\em Keynesian beauty contest} \cite{keynes1937general, marx2022keynesian} is perhaps one of the most essential game-theoretic concepts for blockchain, representing the case where players vote based on the prediction of other players’ beliefs rather than their own.\footnote{Recently, the concept of Keynesian beauty contest has been generalized as the {\em p-Beauty contest game} \cite{moulin1986game, nagel1995unraveling}. This is a number-guessing game where players predict mean value of submitted numbers, multiplied by $p \in (0, 1]$. If the game involves guessing the mean value (i.e., $p = 1$), there exists as many Nash equilibria as the number of choices.}
The Keynesian beauty contest could occur in the Bitcoin protocol and the token-staking (not voting) rule.
Peers who want to maximize their expected rewards might decide on the block to connect \cite{ethereumEpsilonAttack} or the choice to stake their tokens \cite{asgaonkar2018token, wang2019enhancing, tsoukalas2020token} based on the prediction of other peers’ beliefs rather than their own.
This case presents a challenge in eliciting true beliefs from each peer.
Nevertheless, the Bitcoin protocol and the token-staking rule remain popular for two reasons.
First, experimental studies \cite{mehta1994nature} and behavioral game theory \cite{schelling1980strategy} have shown the existence of unique Nash equilibria ({\em Schelling points}) even in the Keynesian beauty contest, suggesting that it is critical in theory but less so in practice.
Second, as shown in \ref{spam} and \ref{free}, these consensus-building approaches help address other obstacles, such as spamming, Sybil attacks, and free-riding.

\subsection{How to Address Spamming and Sybil Attack} \label{spam}
Moreover, consensus-building needs to address {\em spamming} \cite{sahami1998bayesian, hayati2010definition} and {\em Sybil attacks} \cite{douceur2002sybil},
where the former means ``the act of spreading unsolicited and unrelated content \cite{hayati2010definition}," and the latter means ``the forging of multiple identities \cite{douceur2002sybil}."
For example, peers in The DAO might create numerous meaningless proposals to disrupt consensus-building (spamming) or pretend to be different individuals even though they are controlled by one entity (Sybil attack).
A decentralized system must address these problems without relying on the management of some centralized entity. 

Blockchain-related products have leveraged the traditional solution for spamming, assigning a small cost for each transaction.
This method, initially proposed as {\em Hashcash} \cite{back2002hashcash}, adds a minor computational cost to sending email.
The cost accumulates as the number of transaction increases, making spamming unprofitable.
Protocols like Bitcoin and Ethereum require {\em transaction fees} in Bitcoin or Ether, which is paid by the transaction sender to the miner (validator) who includes the transaction in a valid block.\footnote{Note that Hashcash is also the origin of proof-of-work. The Bitcoin protocol has applied the solution for spamming to the solution for strategic behavior in consensus-building as well.}
DApps usually impose additional fees in tokens whenever peers create a proposal that requires consensus-building \cite{nounsNouns, uniswapIntroducing}.

For Sybil attacks, blockchain-related products ususally address this problem by making voting power independent of individuals.
In the Bitcoin protocol, the probability of creating a new block is proportional to computational resources (proof-of-work), i.e., one CPU = one vote.
The token-staking has similar objective; voting power is (in most cases) proportional to the token amount, i.e., one token = one vote.
This approach would be natural in an environment where individuals can make any number of peers.
Alternatively, some platforms, like {\em Gitcoin} \cite{a2017_gitcoin} (see also \ref{casegit}) and {\em Worldcoin} \cite{worldcoin}, use more direct methods like requiring social media accounts or employing biometric authentication using the iris.
While these models are not as decentralized as the one CPU = one vote or the one token = one vote models, they offer more direct ways of establishing individual identities, enhancing the efficiency of consensus-building.\footnote{See also {\em decentralized identifiers} (DIDs) \cite{avellaneda2019decentralized, reed2020decentralized} as a similar concept.} 

Prior studies in economics (in addition to axiomatic ones \cite{chen2019axiomatic, leshno2020bitcoin}) have also addressed spamming and Sybil attacks.
Spamming has been modeled as a game between a spammer and a detector \cite{androutsopoulos2005game}, including models that impose a small cost on each transaction, similar to blockchain solutions \cite{reshef2006effects}.\footnote{See Rao and Reiley (2012) \cite{rao2012economics} for other economic analysis on spamming.}
Sybil attacks have also been examined from a game-theoretic perspective, with studies \cite{gatti2004sufficiently, margolin2007informant, kumar2020game} proposing various mechanisms involving rewards, penalties, and costs for mitigation.
Such game-theoretic approaches to Sybil mitigation have recently been applied to blockchain-related products, especially to the protocol alternative to proof-of-work and proof-of-stake \cite{samanta2023game, stodt2023introducing}.\footnote{See Levine et al. (2006) \cite{levine2006survey} for other ways to mitigate Sybil attacks.}

\subsection{How to Address Free-riding Problem} \label{free}
Even if we can prevent strategic behavior, spamming and Sybil attack, how can we facilitate consensus-building participation?
Given the time and effort for assessment, peers may not commit to consensus-building in the first place or provide uninformative reports independent of true beliefs (e.g., automatically sending the same report). 
Such behavior, described as ``an individual user who uses the system resources without contributing anything to the system \cite{ramaswamy2003free}" is typical in peer-to-peer systems as a {\em free-riding problem}.\footnote{Note that in economics, the term 'free-riding' is specifically used for non-excludable goods and carries a slightly different meaning than it might in the blockchain context \cite{hardin2003free}.}
 
Blockchain-related products have addressed this free-riding problem simply by rewarding participants.
The Bitcoin protocol provides newly minted Bitcoin (i.e., coinbase) to the peer who creates a block in the longest chain.
Token-staking usually involves redistributing staked tokens (e.g., The DAO) or issuing new tokens to peers participating in consensus-building (e.g., Ethereum);
however, these rewards as counter-measures can induce strategic behaviors such as the Keynesian beauty contest.\footnote{According to Appendix A of Ito (2021) \cite{ito2021consensus}, the simple rule of redistributing staked tokens in token-staking rules does not seem to increase the expected amount of rewards. Therefore, issuing new tokens is necessary to effectively address the free-riding."}

\begin{figure}[t]
\centering
 \includegraphics[width=1.0\hsize]{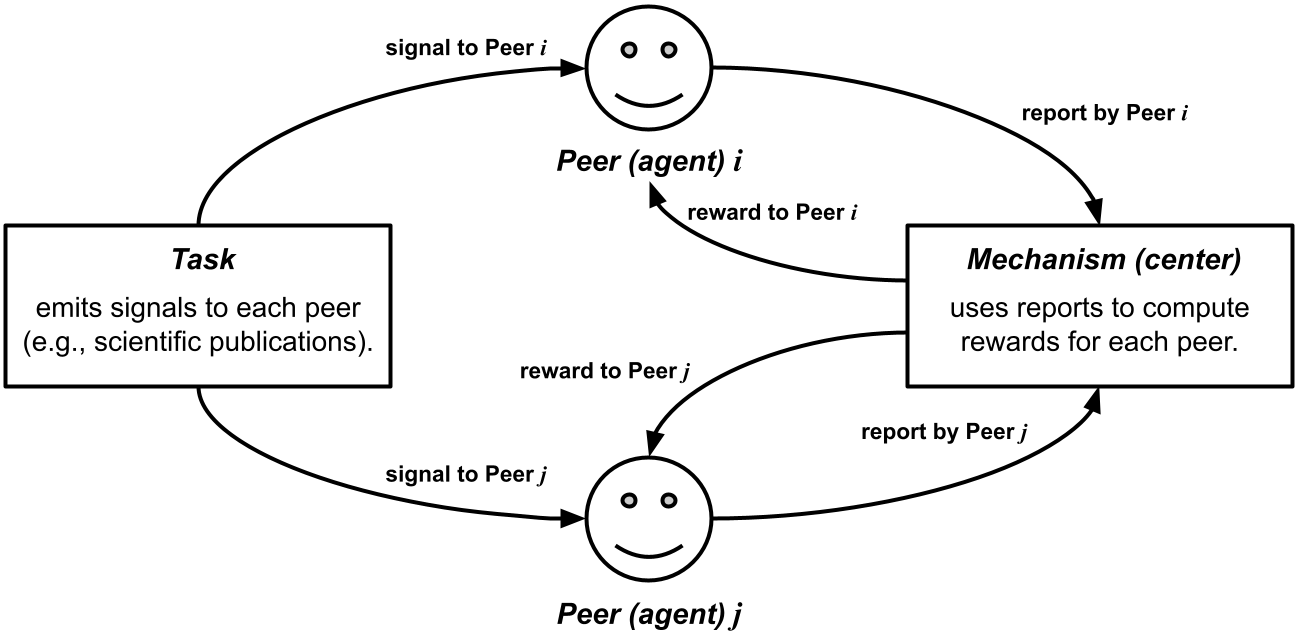}
 \caption{Mechanisms for Information Elicitation}
 \label{pp}
\end{figure}

Few prior studies in economics cover free-riding because game theory or mechanism design implicitly assumes that players participate in consensus-building without any rewards (e.g., auction, voting);\footnote{To the author's knowledge, Thum (2018) \cite{thum2018economic} and Soria (2020) \cite{soria2020tullock} have analyzed whether it is reasonable for peers to participate in the consensus-building of the Bitcoin protocol, using the framework of {\em Tullock contests} \cite{tullock1967welfare}.}
however, this topic has been studied in DAMD, especially in fields applying game theory to elicit true beliefs from peers (agents) in systems such as crowdsourcing, rating systems, and federated learning \cite{faltings2017game, faltings_2023_gametheoretic}.\footnote{This topic is also known as {\em information elicitation without verification} (IEWV).}
These studies generally assume that peers i) first report the content of stochastic {\em signals} emitted from the assigned {\em tasks} ii) and then obtain {\em rewards} whose amount is computed from collected reports (Figure \ref{pp}).
The goal is to design a mechanism that provides maximum expected rewards when peers report accurate signals.
This field has proposed various mechanisms such as {\em Bayesian truth serum} (BTS) \cite{prelec2004bayesian}, {\em peer-prediction} \cite{miller2005eliciting}, and {\em correlated agreement} (CA)\cite{dasgupta2013crowdsourced}. 
For example, the following is an overview of a simple CA mechanism designed by Dasgupta and Ghosh (2013) \cite{dasgupta2013crowdsourced} (DG13):

\begin{tcolorbox} [title={\em CA Mechanism by Dasgupta and Ghosh (DG13)}]

DG13 considers the following situation:

\begin{itemize}
    \item Two peers, $i$ and $j$, each review multiple tasks.
    \item They report binary signals, e.g., \{accept, reject\}.
    \item Each time $i$ and $j$ review the same task, they receive a reward $x$ calculated as follows:
\end{itemize}
    
\begin{equation*} 
  x = \delta (r_{i}, r_{j}) - \delta (r_{i}^{\prime},r_{j}^{\prime}),
\end{equation*}

\noindent    
where $r_{i}$ and $r_{j}$ are the reports of $i$ and $j$ for the same task;
$r_{i}^{\prime}$ and $r_{j}^{\prime}$ are other reports of $i$ and $j$, selected randomly from their previous ones.
$\delta$ is the Kronecker delta denoting the following function:
    \begin{align}
        \delta(a,b) \ =\
        \begin{cases}
            0 & {\rm if}\;\; a \neq b, \\
            1 & {\rm if}\;\; a = b.
        \end{cases} \nonumber
    \end{align}

DG13 has two assumptions:
\begin{itemize}
    \item Tasks emit binary signals, e.g., \{accept, reject\}, positively correlated among tasks. 
    \item Peers take a mixed strategy for received signals.\footnote{Namely, peers decide $p_1 \in \lbrack0,1\rbrack$ and $p_2 \in \lbrack0,1\rbrack$, where they report {\em accept} with probability $p_1$ and {\em reject} with probability $(1 - p_1)$ if an {\em accept} signal is received. Peers report {\em accept} with probability $p_2$ and {\em reject} with probability $(1 - p_2)$ if a {\em reject} signal is received.}
\end{itemize}

\end{tcolorbox}


\noindent
Despite its simplicity, this mechanism achieves the solution concept of {\em strongly truthfulness} \cite{dasgupta2013crowdsourced} where peers maximize their expected rewards by constantly reporting accurate signals or constantly reporting wrong signals.\footnote{Shnayder et al. (2016) \cite{shnayder2016informed} has extended DG13 from binary to multiple signals.}

For task allocation and reward computation, mechanisms on information elicitation typically assume the presence of centralized authorities (Figure \ref{pp}); however, several studies \cite{goel2020peer, goel2021infochain, moti2020orthos, ito2019token, ito2021consensus} have attempted to apply these mechanisms to decentralized blockchain-related products.
This includes efforts to integrate information elicitation with information aggregation, aiming to derive consensus from elicited information \cite{cai2020truth}.

\subsection{How to Address Other Norms} \label{norm}

Considering the obstacles to designing consensus-building also involves considering the norms of what constitutes `good' design.
Consensus-building can be `good' by addressing issues like strategic behavior, spamming, Sybil attacks, and free-riding.
Furthermore, the blockchain context has fostered a variety of other norms, especially in the process of improving the Bitcoin protocol.

One of the most critical norms is {\em scalability} \cite{bondi2000characteristics}.
Decentralized consensus-building faces congestion issues with the increase in transactions and participants, separate from spamming and Sybil attacks.
Compared to centralized systems, decentralized protocols typically have lower {\em transaction per second} (TPS) (Table \ref{tps}),\footnote{According to official statements \cite{visa, visaVisaCrypto}, VISA's TPS has a capacity exceeding 65,000. The figure of 1,700 TPS mentioned in the table is an estimated based on its current usage.} presenting a major hurdle for their adoption in peer-to-peer electronic cash systems or DApp platforms. 
To address this, blockchain-related products often resort to the {\em Layer 2} approach, which involves creating an additional protocol layer to aggregate and process transactions more efficiently. 
Examples include the Lightning Network \cite{poon2015bitcoin} and {\em ZeroSync} \cite{linuszerosync} for the Bitcoin protocol, and {\em Arbitrum} \cite{youtubeArbitrumBlockchainbased, kalodner2018arbitrum} and {\em ZK-SNARKS} \cite{ben2014succinct}\footnote{ZK-SNARKS employs a cryptographic method known as {\em zero-knowledge proof} (ZKP) \cite{quisquater1989explain}. In the blockchain context, ZKP was initially used for privacy-focused initiatives, such as anonymous money transfers \cite{sun2021survey}. Its applications to scalability issues were later promoted by Buterin \cite{vitalikIncompleteGuide, vitalikApproximateIntroduction}.} for Ethereum.
In this context, incentive design and cryptography play vital roles in preserving decentralization at the Layer 2 level.\footnote{Some argue that blockchain faces a {\em trilemma} \cite{9031204}, suggesting that it can achieve only two out of the three goals: scalability, decentralization, and security. This concept was recently formulated by Nakai et al. (2023) \cite{nakai2023tri}.}
The extant literature provides details of blockchain scalability \cite{sanka2021systematic, thibault2022blockchain}.

\begin{table}[t]
 \caption{Transaction per Second (TPS) as of October 2023}
 \label{tps}
 \begin{center}
 \rowcolors{1}{white}{gray!10}
  \begin{tabular}{ll}
   \toprule
   \textbf{Protocols} & \textbf{approx. TPS} \\[2pt]
   \midrule
{\em Bitcoin} \cite{nakamoto2008bitcoin} & $5$ - $7$ \\[2pt]
{\em Ethereum} \cite{buterin2014ethereum, wood2014ethereum} & $12$ - $15$ \\[2pt]
{\em VISA} & $1,700$ \\[2pt]
   \bottomrule
  \end{tabular}
 \end{center}
   \captionsetup{font=footnotesize, justification=raggedright, singlelinecheck=off}
    \caption*{{\em Sources}: 
    Blockchain.com (\href{https://www.blockchain.com/explorer/charts/transactions-per-second}{https://www.blockchain.com/explorer/charts/trans\\actions-per-second}, accessed September 4, 2023), 
    Dune.com (\href{https://dune.com/k06a/TPS}{https://dune.com/k06a/TPS}, accessed September 4, 2023), 
    HackerNoon.com (\href{https://hackernoon.com/the-blockchain-scalability-problem-the-race-for-visa-like-transaction-speed-5cce48f9d44}{https://hackernoon.com/the-blockchain-scalability-problem-the-race-for-visa-like-transaction-speed-5cce48f9d44}, accessed September 4, 2023)}
\end{table}

Other important norms for consensus-building include {\em finality} \cite{cointelegraphWhatFinality} and {\em energy efficiency} \cite{de2018bitcoin}.
Nakamoto consensus, used in the Bitcoin protocol, cannot guarantee finality (the state that will not be reverted) because the longest chain is probabilistic, not deterministic, and proof-of-work requires significant energy for consensus.\footnote{See the Hashrate chart \cite{bitinfochartsBitcoinHashrate} for information on the computational resources used by the Bitcoin protocol for consensus-building. Some studies criticize this level of resource consumption \cite{stoll2019carbon}, while others defend it \cite{khazzaka2022bitcoin}.}
In contrast, consensus-building based on token-staking are preferred for their determinism and greater energy efficiency. 
Regarding energy efficiency, blockchain-related products are exploring the use of proof-of-work computational resources for additional purposes, such as prime number discovery \cite{king2013primecoin} and optimization problem-solving \cite{shibata2019proof}.

Moreover, several products have introduced new norms for determining peers' voting power. 
In decentralized storage networks \cite{benisi2020blockchain, khalid2023comprehensive} (e.g., {\em Filecoin} \cite{benet2018filecoin}, {\em Arweave} \cite{williams2019arweave}, and {\em Sia} \cite{vorick2014sia}), the voting power is allocated based on the amount of storage a peer contributes. 
{\em NEM} \cite{nem2018nem} has implemented {\em proof-of-importance}, which combines elements of proof-of-work, proof-of-stake, and the network structure of token transactions.
Gitcoin \cite{a2017_gitcoin} adopts {\em quadratic voting} (QV) \cite{posner2015voting, weyl2017robustness, lalley2018quadratic, buterin2019flexible}, a derivative of token-voting, where the voting power of a peer for an option is the square root of the token amounts they have staked in that option.\footnote{It is important to note that alternative models, which depart from the one CPU = one vote or one token = one vote paradigms, might be more prone to strategic behaviors, spamming, Sybil attacks, and free-riding. Specifically, QV can be vulnerable to collusion and Sybil attacks, as explored in Section \ref{casegit}.}
The extant literature provides the details of blockchain consensus-building \cite{cachin2017blockchain, bano2019sok, ferdous2020blockchain, ding2023voting}.\footnote{Prior studies in economics, particularly in social choice theory \cite{arrow2012social}, have also fostered norms related to consensus-building. These norms serve as the foundation for game theory and mechanism design. Comparing the norms of economics with those of blockchain, or analyzing discussions by interchanging these norms, could serve as another approach to bridge the contexts of both fields.}


\vspace{0.4\baselineskip}
This section surveyed how blockchain-related products and prior studies have addressed the obstacles of designing consensus-building for decentralization, where obstacles include strategic behavior, spamming, Sybil attacks, and free-riding.
A practical design should address these obstacles simultaneously.

\section{Designing Token Value for Autonomy} \label{token}

The previous discussion implicitly assumes that peers act to maximize the amount of their expected rewards;
however, if rewards are in tokens, their value and sufficiency as incentives become critical. For instance, the behavioral response of peers would vary significantly if the reward tokens were exchangeable for 0.1 US dollars (USD) compared to 1,000 USD. 
This section surveys how blockchain-related products and prior studies have addressed the design of token value for autonomy, employing a microeconomic framework.

\subsection{How to Ensure Market Price (Supply Side)} \label{tokensupply}

To be considered valuable, tokens must ensure their market price.
As mentioned in \ref{oppre}, the market price of goods is established at the intersection of the supply-side value on the marginal cost and the demand-side value on the marginal utility \cite{marshall1890principles}.

On the supply side, the value depends on the marginal cost, i.e., the cost added by producing one additional unit of a product or service. 
Intuitively, we continue to produce goods as long as we can sell them at a price above their marginal cost.

Blockchain-related products often inherently embody this principle. 
For example, the Bitcoin protocol provides a coinbase for the peer who succeeds in making a new valid block at the expense of proof-of-work, meaning creating one additional Bitcoin unit requires computational resources.
Moreover, the following additional rules are worth noting for the marginal cost of Bitcoin:

\begin{tcolorbox} [title={\em Additional Rules of the Bitcoin Protocol}] 
    \begin{itemize}
        \item The amount of coinbase halves for every 210,000 blocks; this halving continues until all 21 million Bitcoins are mined ({\em block-reward halving}).\footnote{\url{https://www.bitcoinblockhalf.com/}}
        \item The difficulty of proof-of-work changes for every 2,016 blocks to keep the block interval 10 minutes ({\em difficulty adjustment}).\footnote{\url{https://www.blockchain.com/explorer/charts/difficulty}}
    \end{itemize} 
\end{tcolorbox}

\noindent
The extant literature provides more details \cite{nakamoto2008bitcoin, bitcoinControlledSupply, bitcoinDifficultyBitcoin}.
Notably, the former {\em block-reward halving} works to gradually increase the marginal cost.\footnote{It is also worth noting that block-reward halving contributes to user acquisition in the early phase. This is because it offers a kind of first-mover advantage; specifically, peers who participate in consensus-building early can earn more Bitcoin as a reward.}
In other products, marginal costs extend beyond computational resources. 
Ethereum uses the opportunity cost of staking because its proof-of-stake requires token staking to participate in consensus-building.
Decentralized storage networks (\ref{norm}) use storage space, while
The DAO (and many other DApps) uses human resources (in addition to the opportunity cost of staking) because its consensus-building deals with subjective issues.

The marginal cost (of a token) may depend on another token.
For example, many DApps adopt {\em initial coin offering} (ICO) for initial token distribution \cite{kher2021blockchain, magnusson2022initial}, where we can obtain tokens in exchange for other tokens (usually Ether) at a given rate for a given period.\footnote{Ethereum itself was also launched through an ICO using Bitcoin \cite{ethereumLaunchingEther}, even though the term ICO was not established. To the best of the author's knowledge, the first ICO using Ether (termed a crowdsale then) was conducted by The DAO. Both ICOs incentivized early participation by gradually increasing the exchange rate during the sale period.}
In this case, marginal cost (at initial distribution) is linked to the value of another token.
Some DApps (e.g., {\em Edgeware} \cite{edgewareSmartContract} and {\em Astar} \cite{astarPlasmLockdrop}) have utilized {\em lockdrops} for initial token distribution, where we can obtain tokens by depositing another token (usually Ether) for a given period.
Unlike ICOs, locked tokens will be returned after the period (e.g., one year).
In this case, marginal cost is linked to the opportunity cost of another token.\footnote{Another distribution method is the {\em airdrop}, where tokens are freely distributed to peers (accounts) meeting specific criteria (e.g., having used the DApp before). Although this method appears to incur little marginal cost, the rationale behind airdrops has been analytically explored by Allen et al. (2023) \cite{allen2023airdrop}.}

Prior studies in economics focused on token valuation \cite{liu2022crypto} rather than design, potentially because economics has already concluded that we can design supply-side value by imposing some cost on new token issuance (and it is even better if that cost increases over time).
Marginal cost serves as a practical metric for valuation, particularly for Bitcoin, where computational resources are quantifiable.
Hayes (2015, 2017, 2019) \cite{hayes2015cost, hayes2017cryptocurrency, hayes2019bitcoin} developed a cost-of-production model for Bitcoin valuation,\footnote{He has developed this model as a counter to the Yermack (2015) \cite{yermack2015bitcoin}'s argument that Bitcoin has no intrinsic value.} which has extended to address halving effect\footnote{Pagnotta and Buraschi (2018) \cite{pagnotta2018equilibrium} argues that the block reward-halving may have both positive and negative effect on the Bitcoin price, as it reduces the increasing rate of total supply, not the total supply itself (i.e., disinflation, not deflation).} \cite{pagnotta2018equilibrium, meynkhard2019fair} and analyze other tokens \cite{hayes2017cryptocurrency}.
For proof-of-stake, Fanti et al. (2019) \cite{fanti2019economics} modeled the opportunity cost of staking as the expected rate of return on a risk-matched investment strategy in financial markets.

\subsection{How to Ensure Market Price (Demand Side)} \label{tokendemand} 

On the demand side, the value depends on the marginal utility, i.e., the benefit gained from consuming one additional unit of a product or service.
Intuitively, we continue to consume goods as long as we can buy them at a price below their marginal utility. 

\begin{figure}[t]
\centering
 \includegraphics[width=0.9\hsize]{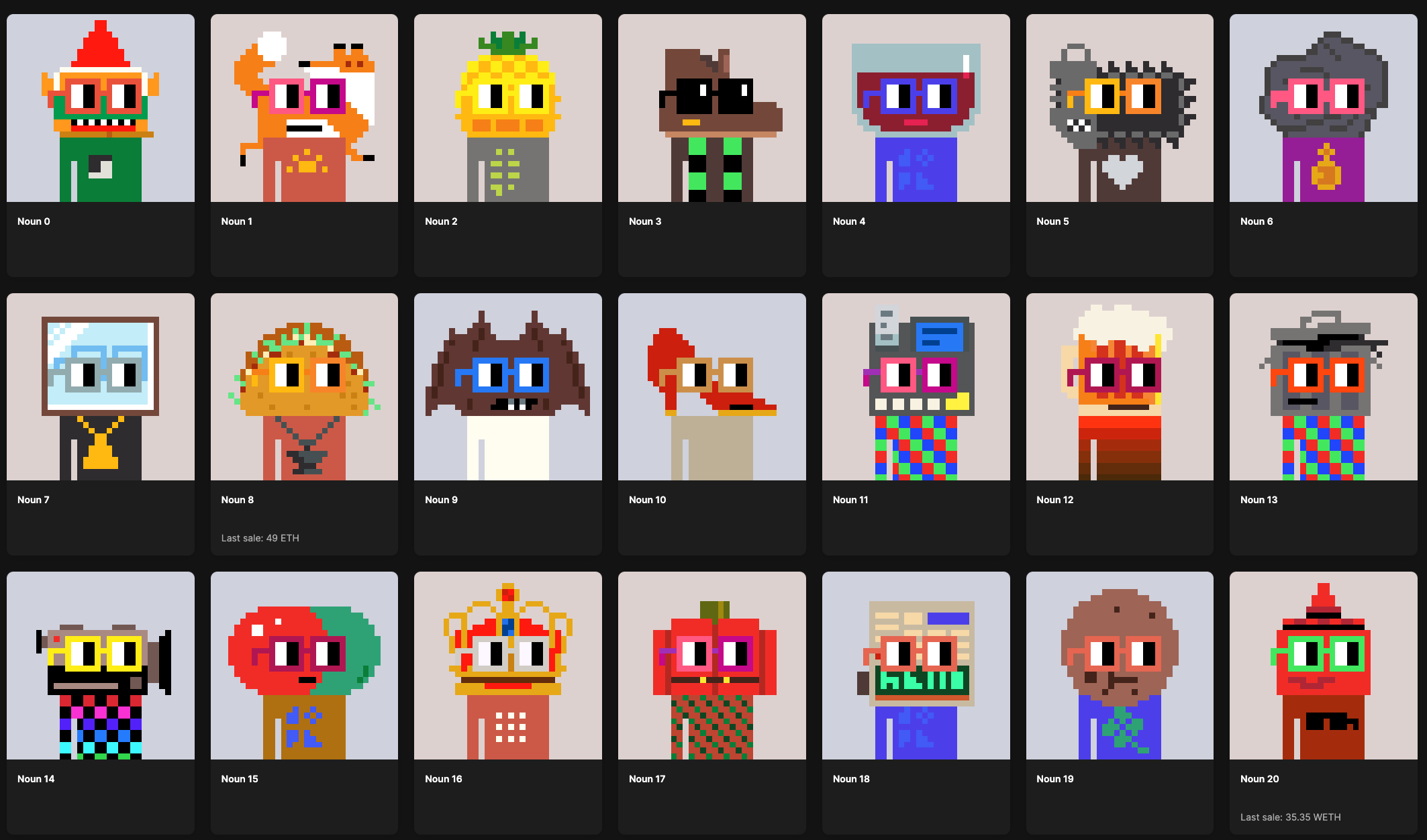}
 \caption{NFTs of Nouns DAO}
 \captionsetup{font=footnotesize, justification=raggedright, singlelinecheck=off}
    \caption*{{\em Source}: OpenSea.io (\url{https://opensea.io/collection/nouns?search[collections][0]=nouns&search[sortBy]=CREATED_DATE}, accessed November 8, 2023).}
    \captionsetup{font=normal}
 \label{nounsfig}
\end{figure} 

Blockchain-related products have designed tokens with diverse marginal utilities.
For example, the Bitcoin protocol provides utility as an electronic peer-to-peer cash system, where its marginal utility may exponentially increase with the number of users ({\em Metcalfe's law} \cite{keymet}).
Bitcoin also has the utility of a transaction fee in the protocol.
Ether has the utility of being able to use DApps on Ethereum; 
this is why Ether is often called as digital-oil while Bitcoin is called as digital-gold \cite{coinbaseWhatEthereum, goldoil, keygoldoil}.

DApps have further diversified marginal utility of tokens.
A common approach is the analogy with stocks, i.e., token holders can gain governance rights and profit shares.  
Such tokens are typical in DAOs and {\em decentralized exchanges} (DEXs) \cite{lehar2021decentralized, xu2023sok}, a DApp for exchanging tokens without a centralized entity (e.g., {\em Uniswap} \cite{uniswapHomeUniswap}, {\em Curve} \cite{curve}).
DEXs usually create a {\em liquidity pool} for exchange, a vault of two (or more types of) tokens collected from peers (see \ref{tokenpeg} for details). 
Here, peers who provided tokens to the pool can earn {\em liquidity provider (LP) tokens} that reward holders with a portion of the fees from DEX users.
DEXs may also issue {\em governance tokens} (e.g., UNI in Uniswap, CRV in Curve) that allow holders to participate in voting on system updates (see \ref{caseuni} for details).
Another notable approach is the analogy with art, i.e., token holders can enjoy the design and rarity of tokens. 
Such tokens are referred to as {\em non-fungible tokens} (NFTs) \cite{ante2022non, ko2023survey}, i.e., a standard to make the token a unique, collectible item (e.g., {\em CryptoKitties} \cite{cryptokittiesCryptoKittiesCollect} and {\em CryptoPunks} \cite{larvalabsCryptoPunks}).
NFTs have an advantage over existing art in that, once issued, their authenticity and provenance are guaranteed (as they are on-chain data). 
Note that these approaches can be combined.
For example, Nouns DAO \cite{nounsNouns} issues NFTs every once a day as collectible items (Figure \ref{nounsfig}), which can be used as governance tokens as well (see \ref{casenouns} for details).
The extant literature provides a detailed classification of token utility \cite{oliveira2018token, lo2020assets, eshraghi2023approaches}.

Prior studies in economics have used various proxies to quantify marginal utility.
For the Bitcoin protocol, the number of users has been used as a standard proxy \cite{peterson2018metcalfe, van2018alternative}.
For more general tokens, Cong et al. (2021) \cite{cong2021tokenomics} developed a dynamic valuation model where tokens' current and future utilities are derived from the number of users and transactions. 
Liu (2022) \cite{liu2022crypto} proposed a model that estimates token utility with token velocity, staking ratio, and the dilution rate.\footnote{Specifically, token velocity is ``the percentage of tokens transacted over a specific period relative to the token supply \cite{liu2022crypto};" staking ratio is ``the proportion of staked tokens to the total token supply \cite{liu2022crypto};" the dilution rate is ``the annual growth rate of the token supply \cite{liu2022crypto}."}  
For tokens with regular income (e.g., LP tokens), traditional asset valuation models, such as the discounted cash flow method, can be applied \cite{brucker2022fi}.
NFT valuation considers these proxies \cite{kraussl2022non} and unique attributes like visual images \cite{nadini2021mapping} and descriptions \cite{horky2022price}.\footnote{{\em Sentiment analysis}, which estimates demand using tools like Google Trends and posts from X (formerly Twitter), is worth mentioning in this context. While sentiment analysis can be applied regardless of the token type, previous studies have primarily focused on short-term price fluctuations \cite{lamon2017cryptocurrency, abraham2018cryptocurrency}. However, for example, Silberholz and Wu (2021) \cite{silberholz2021measuring} attempts to decompose the results of sentiment analysis into speculative and utility-related elements.}

\subsection{How to Stabilize Market Price} \label{tokenstabilize}
Given its role as an incentive, the token's market price should not fluctuate too volatilely.
Market prices fluctuate due to speculative activities and unpredictable events, which may seem difficult to control by design.
Nevertheless, protocols like Bitcoin and Ethereum include stabilizers that indirectly mitigate price fluctuations.

For example, the Bitcoin protocol includes the difficulty adjustment mentioned above that increases proof-of-work difficulty when the final 2,016 blocks are generated too quickly, and vice versa.
Difficulty adjustment, even intended for constant block-interval, can contribute to the price stabilization in terms of making marginal cost predictable and less volatile, i.e., a sudden increase in computational resources may subside with the next difficulty increase.\footnote{Kj\ae rland et al. (2018) \cite{kjaerland2018analysis} empirically observed that Bitcoin difficulty adjustment does not affect price fluctuation.}
An auction-based transaction fee is another stabilizer for the Bitcoin protocol, where senders add a voluntary fee to their transactions, and transactions are stored in blocks in order of the highest amount (i.e., first-price auction). 
In contrast, Ethereum recently switched to the combination of auction-based and posted-price fees \cite{ethereumEIP1559Market}, where the latter amount is set according to the size of the previous block.
These fee mechanisms can stabilize prices to reduce network congestion by making marginal utility (concerning remittances) less volatile.  




Prior studies in economics have analyzed and designed these stabilizers.
Saito and Iwamura (2019) \cite{saito2019make} and Iwamura et al. (2019) \cite{iwamura2019can} present price-stabilization proposals for the Bitcoin protocol, which include performing the difficulty adjustment only when the block interval exceeds a certain threshold.\footnote{Other proposals are i) making the amount of mining rewards variable, dependent on the results of difficulty adjustment (instead of block-reward halving) and ii) introducing a negative interest rate (depreciation) on all Bitcoins.}
In contrast, Noda, et al. (2020) \cite{noda2020economic} indicates the advantages of more frequent difficulty adjustments (e.g., {\em Bitcoin Cash} adjusts the difficulty of every block instead of 2016 blocks).
Fee mechanisms are mainly studied by auction theory, and prior studies have recommended introducing a second-price auction \cite{basu2019towards} or monopolistic auction (collecting the same amount of fees from each transaction in a block) \cite{lavi2022redesigning, yao2018incentive} for the Bitcoin protocol.
Ethereum's fee mechanism has also been analyzed \cite{roughgarden2020transaction, roughgarden2021transaction},
and design proposals include using bids and the size of previous blocks to calculate posted-price fees \cite{ferreira2021dynamic} and combining second-price auction and burning mechanism \cite{chung2023foundations}.


\subsection{How to Stabilize Market Price (Pegging)} \label{tokenpeg}

A more direct approach to price stabilization is to pre-define the exchange ratio between a token and another asset.

The simplest scheme is to peg the value of a token with another asset (e.g., one token = 1 USD); such tokens are known as {\em stablecoins}.
Stablecoins are often issued by a centralized custodian holding the underlying assets as collateral (e.g., {\em USDT} \cite{tetherTransparency}, {\em PAX Gold} \cite{paxosGoldSafest}).
In contrast, some stablecoins attempt to maintain their peg while ensuring decentralized properties.
For example, {\em DAI stablecoin} \cite{makerdaoMakerDAOUnbiased} is pegged to the USD but collateralized by Ether, where any peer can be a custodian.\footnote{This model necessitates overcollateralization to counter the risk of price volatility in the collateral assets. For instance, to issue one unit of the DAI stablecoin pegged at 1 USD, 1.5 USD worth of Ether as collateral is currently required.}
{\em Basis} \cite{al2017basis} and {\em TerraUSD} \cite{kereiakes2019terra} are products classified as {\em algorithmic stablecoins}, which have no collateral and aim to achieve price stability through algorithmic mechanisms.
However, these projects have not succeeded, as exemplified by the collapse of TerraUSD (see \ref{caseterra} for details), leading to the current absence of effective algorithmic stablecoins.
The extant literature provides more details of stablecoin \cite{ito2020stablecoin, catalini2022some}.

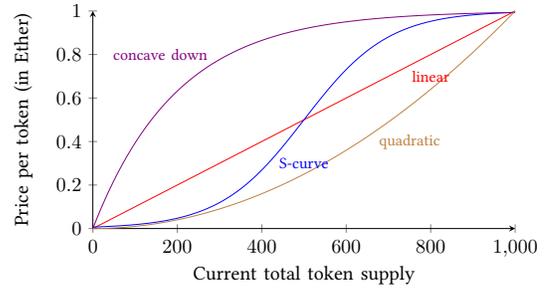
\begin{figure}[t]
\centering
    \begin{tikzpicture}[scale=0.75] 
    \begin{axis}
    [
        axis lines = left,
        xlabel={Current total token supply}, 
        ylabel={Price per token (in Ether)}, 
        xmin=0, 
        xmax=1000, 
        ymin=0, 
        ymax=1,
        width=0.5\textwidth,
        height=0.3\textwidth,
    ]
    \addplot[no marks, domain=0:1000, samples=100, color=red,] {x/1000};
    \addplot[domain=0:1000, samples=100, color=blue,] {1 / (1 + exp(-0.01*x + 5))};
    \addplot[no marks, domain=0:1000, samples=100, color=violet,] {1 - exp(-0.005*x)};
    \addplot[domain=0:1000, samples=100, color=brown,] {x^(2)*(1/1000000)};
    
    \node[color=red, font=\footnotesize] at (axis cs: 800,0.7) {linear};
    \node[color=blue, font=\footnotesize] at (axis cs: 500,0.3) {S-curve};
    \node[color=violet, font=\footnotesize] at (axis cs: 160,0.8) {concave down};
    \node[color=brown, font=\footnotesize] at (axis cs: 750,0.4) {quadratic};
    
    \end{axis}
    \end{tikzpicture}
 \caption{Token Bonding Curve (TBC)}
 \label{bonding}
\end{figure}

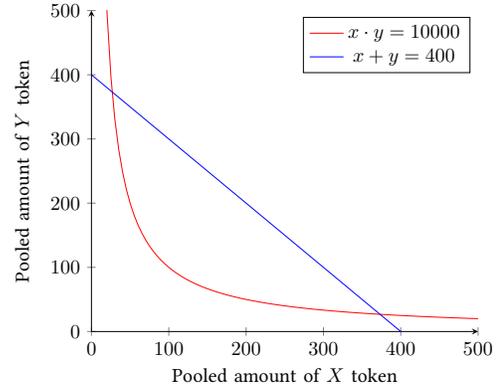
\begin{figure}[t]
\centering
    \begin{tikzpicture}[scale=0.75]
    \begin{axis}
    [
        axis lines = left,
        xlabel={Pooled amount of $X$ token}, 
        ylabel={Pooled amount of $Y$ token}, 
        xmin=0, 
        xmax=500, 
        ymin=0, 
        ymax=500,
    ]
    \addplot[no marks, domain=0:500, samples=100, color=red,] {10000 / x};
    \addlegendentry{$x \cdot y = 10000$}

    \addplot[no marks, domain=0:500, samples=100, color=blue,] {400 - x};
    \addlegendentry{$x + y = 400$}
    
    \end{axis}
    \end{tikzpicture} 
 \caption{Automated Market Maker (AMM)}
 \label{amm}
\end{figure} 

Another scheme is to pre-define rules for exchange ratios.
{\em Token bonding curve} (TBC), one of the first such rules for blockchain, was proposed \cite{Tokens20, vitalikPathIndependence, yosBondingCurves} and implemented \cite{hertzog2017bancor} in 2017.
TBC is a program that mints new tokens by depositing Ether and returns Ether by depositing (burning) the minted tokens, with the exchange ratio (i.e., the price per token in Ether) determined as a function of the current token supply.
This function allows us to exchange tokens without counterparties; in the case of increasing functions (Figure \ref{bonding}), we can also earn Ether by first acquiring tokens and then burning them back to Ether once the supply has increased sufficiently.\footnote{TBC typically employs an increasing function for minting tokens, as decreasing functions provide no initial incentive to mint tokens.}
Moreover, TBC has evolved into a form of {\em automated market maker} (AMM) \cite{zhang2018formal}, which pre-defines the exchange ratio between two tokens as follows:

\begin{tcolorbox} [title={\em Automated Market Maker (a simple example)}]

    Create a liquidity pool:
    \begin{itemize}
        \item Peers can create a liquidity pool comprising two types of tokens, $X$ and $Y$.
        \item A liquidity pool is established when it satisfies the formula:
        \vspace{0.3\baselineskip}
        \begin{equation*}
            x\cdot y = k,
        \end{equation*}
        where $x$ and $y$ denote the pooled amounts of $X$ and $Y$, respectively, and $k$ is a constant value.
    \end{itemize}

    \vspace{0.6\baselineskip}
    
    Exchange tokens in the liquidity pool:
    \begin{itemize}
        \item Peers can exchange $X$ and $Y$ through the liquidity pool.
        \item The exchange ratio maintains the formula above; exchanging $\Delta x$ of $X$ for $Y$ yields $\Delta y$ such that $(x+\Delta x)(y-\Delta y) = k$.\footnote{Exchange fees are omitted for simplicity.}
    \end{itemize}
    
\end{tcolorbox}

\noindent
The red curve in Figure \ref{amm} represents the $k = 10,000$ case, where the price of $X$ (in $Y$) is $1$ for $x = 100$, $y = 100$, and $5$ for $x = 20$, $y = 500$.
This mechanism allows us to exchange tokens without counterparties and provides arbitrage opportunities when price discrepancies arise between the liquidity pool and external markets (i.e., AMM can autonomously reach the appropriate relative price). 
AMM is central to the design of current DEXs and is continually refined through various proposals, including alternative constraint formulas (e.g., $x + y = k$ represented by the blue line in Figure \ref{amm}) and flexible fee structures \cite{adams2021uniswap}.\footnote{The constant product model $x\cdot y = k$ is often called the {\em constant product automated market maker} (CPAMM) to distinguish it from other models. CPAMM is prevalent due to its convenient property that prevents the pool from depleting either $X$ or $Y$.}
The extant literature provides more details \cite{xu2023sok}.

TBC and AMM were designed to increase token liquidity (i.e., to facilitate exchange with other tokens)\footnote{Earlier DEXs like the {\em 0x protocol} \cite{0xPowerfulAPIs} required a counterparty (known as a {\em relayer}) for token exchange, which was a significant barrier to practical usage.} but
they also aid in price stabilization by making the token price more predictable.

Prior studies in economics have informed the design of stablecoins, particularly through dynamic analyses of incentives and expectations \cite{catalini2021economic, d2022can}.
For collateralized stablecoins, the behavior of users \cite{routledge2022currency} and custodians \cite{li2022money, d2022can} has been studied, drawing from theories of currency crises \cite{obstfeld1996models} and bank runs \cite{diamond1983bank};
these studies often emphasize the importance of the custodian's {\em commitment} \cite{kandori2023mighty}.
For algorithmic stablecoins, their sustainability has been studied \cite{fu2023rational} while modeling them as a {\em Ponzi scheme}, reflecting the failure of preceding proposals \cite{sams2015note}.
Studies on TBC and AMM are mainly focused on formalization, e.g., TBC \cite{zargham2020curved, zargham2020economic} and AMM \cite{angeris2020improved, bartoletti2022theory},
and the business side is ahead concerning their design \cite{adams2021uniswap, adams2023uniswap}.\footnote{To the author's knowledge, Krishnamachari et al. (2021) \cite{krishnamachari2021dynamic}, which proposed dynamic constraint formula, is a study on the AMM design.}
Nontheless, room remains for economics to contribute to their design, given that i) AMM is clearly an application of an {\em indifference curve} and {\em marginal rate of substitution} in microeconomics,\footnote{An {\em indifference curve} represents a set of points, each denoting a combination of two goods or services that provide equal utility to the consumer. The {\em marginal rate of substitution} is the rate at which a consumer is willing to exchange one good for another while maintaining the same level of utility, represented by the slope of the indifference curve. See Kandori (2023) \cite{kandori2023mighty} for the detail of these concepts.} and ii) the scoring rules \cite{hanson2003combinatorial, othman2012automated} underlying TBC and AMM have been merged with game theory to produce various models for information elicitation (\ref{free}).

\vspace{0.4\baselineskip}

This section surveyed how blockchain-related products and prior studies have designed token value for autonomy.
From an economic perspective, token value can be ensured by marginal cost and marginal utility, and stabilizers must adjust at least one (or pre-define the exchange ratio of tokens).
Practical design requires imposing both marginal cost and marginal utility on the tokens.

\section{Case Studies} \label{case}
Thus far, we have explored consensus-building for decentralization and token value for autonomy; however,
as mentioned in Section \ref{history}, practical design necessitates their integration.
This section assesses five previously-mentioned products from the viewpoint of integration (i.e., whether they cover multiple requirements simultaneously).
The summary of these evaluations is presented in Table \ref{cases}.

\begin{table*}[t]
 \caption{Evaluation of Five Blockchain-related Products}
 \label{cases}
 \begin{center}
 \rowcolors{1}{white}{gray!10}
 \begin{adjustbox}{center}
  \begin{tabular}{p{4.6cm}cp{2.8cm}p{2.5cm}p{1.6cm}p{1.8cm}p{1.2cm}}
   \toprule
   \multicolumn{1}{c}{\textbf{Products \& Notes}} & \multicolumn{3}{c}{\textbf{Consensus-Building for Decentralization}} & \multicolumn{3}{c}{\textbf{Token Value for Autonomy}} \\[2pt]
     & strategy-proofness & spam \& Sybil-proofness & free-riding proofness & marginal cost & marginal utility & \:stabilizer \\
   \midrule
{\em The Bitcoin Protocol} \cite{nakamoto2008bitcoin} \newline benchmark of integration & \checkmark & \centering \checkmark & \centering \checkmark & \centering \checkmark & \centering \checkmark & \quad\:\:\:\:\checkmark \\[12pt]
{\em Gitcoin} \cite{a2017_gitcoin} \newline QV at the expense of decentralization & & \centering\quad\:\:\:\:\:\checkmark \newline(centralized) & \centering\vspace{0.09cm}(not required) & & \centering\checkmark & \\[12pt]
{\em Nouns DAO} \cite{nounsNouns} \newline room for NFT-specific extensions & \checkmark & \centering \checkmark & & \centering\checkmark & \centering\checkmark &  \\[12pt]
{\em Terra} \cite{kereiakes2019terra} \newline lack of marginal cost led to collapse & \checkmark & \centering \checkmark & \centering\checkmark & & \centering\checkmark & \quad\:\:\:\:\checkmark \newline(collapsed) \\[12pt]
{\em Uniswap} \cite{uniswapHomeUniswap} \newline tokens with unclear raison d'etre & \checkmark & \centering \checkmark & & \centering\quad\:\checkmark \newline(weak) & \centering\quad\:\checkmark \newline(weak) & \\[12pt]
   \bottomrule
  \end{tabular}
  \end{adjustbox}
 \end{center}
\end{table*}

\subsection{The Bitcoin Protocol} \label{casebit}
The Bitcoin protocol \cite{nakamoto2008bitcoin} is a peer-to-peer electronic cash system, the first practical protocol to employ an economic incentive design for consensus-building on transaction records.
We have already confirmed its incentive design but briefly review it again here.

For consensus-building, combining proof-of-work and Nakamoto consensus prevents strategic behavior and free-riding.
Spamming and Sybil attacks are addressed by transaction fees and proof-of-work, respectively.
For token value, the combination of proof-of-work and coinbase stipulates its marginal cost, which increases over time by the block-reward halving and is stabilized (in line with current input computational resources) by the difficulty adjustment.  
Marginal utility is ensured through peer-to-peer electronic cash; transaction fees may also be included. 

Here, a single function has multiple roles (e.g., proof-of-work contributes to preventing strategic behavior and Sybil attacks and ensuring marginal cost), which enables the integration of consensus-building and token value.
Regarding integration, the Bitcoin protocol would be a benchmark for other blockchain-related products.
Further consideration should be given to other norms, such as scalability, finality, energy efficiency, and voting power (\ref{norm}). 
Moreover, {\em governance}\textemdash consensus-building on the protocol itself\textemdash is also an important issue for the Bitcoin protocol, which will be discussed at the end of this survey (\ref{outside}).

\subsection{Gitcoin} \label{casegit}
Gitcoin \cite{a2017_gitcoin} is a DApp on Ethereum, where peers can pool any amount of Ether (or stablecoins) to donate the listed open source projects.
Gitcoin is not entirely DAO (at the time of this writing), as a management team evaluates which proposals should be listed.
Conversely, it adopts a decentralized consensus-building on redistributing pooled assets among listed projects.

\begin{figure}[h]
\centering
 \includegraphics[width=0.9\hsize]{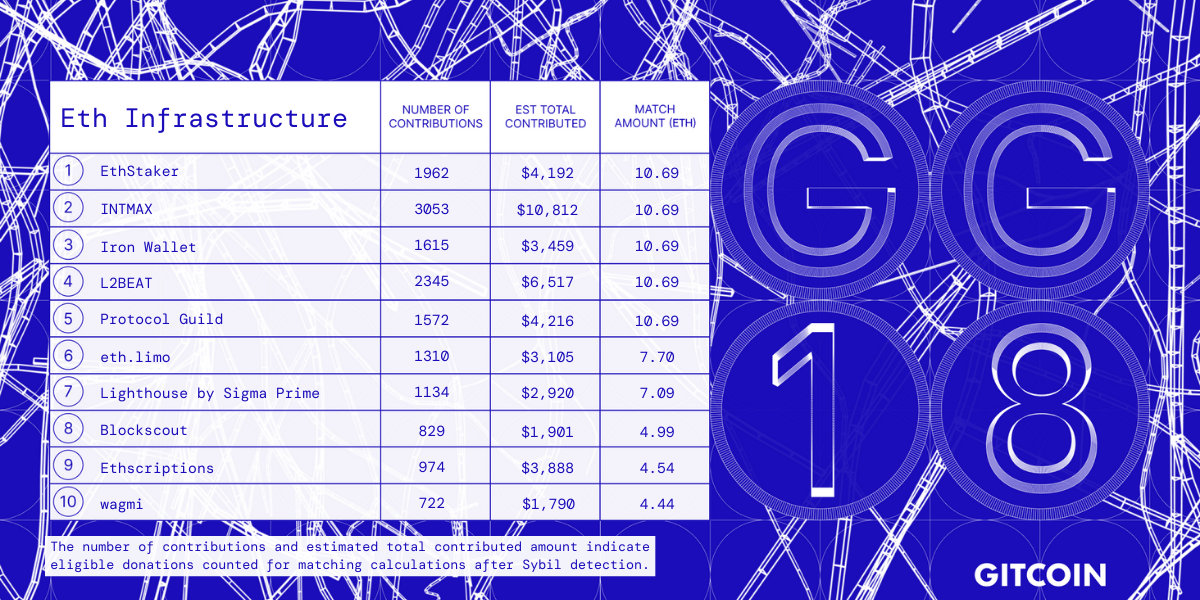}
 \caption{Result of Gitcoin Grants Round 18}
 \captionsetup{font=footnotesize}
    \caption*{{\em Source}: Gitcoin Grants Round 18: Results and Recap (\url{https://www.gitcoin.co/blog/gitcoin-grants-round-18-results-and-recap}, accessed November 8, 2023).}
 \captionsetup{font=normal}
 \label{gitcoin}
\end{figure} 
 
For consensus-building, Gitcoin adopts QV (\ref{norm}) as a counter to the one CPU or one token = one vote norm;\footnote{To be precise, Gitcoin applies QV to fund allocation rather than voting, a method referred to as {\em quadratic funding} \cite{buterin2019flexible}.}
however, QV is vulnerable to collusion (strategic behavior) and Sybil attacks \cite{braun2021incentivization} because the same token amount increases voting power with the number of holders.
Aside from collusion, Sybil attacks are prevented by {\em Gitcoin Passport} \cite{gitcoinWhatGitcoin} in which a management team requires each user to submit one or more social media accounts and pay a certain amount of tokens, guaranteeing as much one-to-one correspondence between addresses and individuals as possible.
Figure \ref{gitcoin} shows some of the results of the latest round of grant programs after QV and Sybil detection (by Gitcoin Passport). 
The management team prevents spamming, and the free-riding proof is not required here because Gitcoin is a platform for donations.
To be DAO, Gitcoin issued Gitcoin Token (GCT) in 2021.\footnote{\url{https://etherscan.io/address/0xde30da39c46104798bb5aa3fe8b9e0e1f348163f}, [Accessed 20-11-2023]}
GCT appears to have no marginal cost for token value as it is a pre-mined token.
Marginal utility relates to Gitcoin Passport and governance.
Peers can increase the identity score of Gitcoin Passport by staking their GCT.
The governance feature is under development, but a QV using GCT will likely be implemented to decide how to improve Gitcoin.
The value of GCT has no stabilizer.

Overall, Gitcoin is a case of introducing a new QV norm at the expense of some decentralization.
The risk of collusion remains even if we can ensure one-to-one correspondence between user addresses and individuals.
Gitcoin tries to make QV and decentralized autonomous property compatible through the issuance of GCT; however, as it lacks a strategy-proof, marginal cost, and stabilizer design, GCT does not contribute much to the purpose (at least for now).

\subsection{Nouns DAO} \label{casenouns}
Nouns DAO \cite{nounsNouns} is another DApp on Ethereum that 
i) automatically generates a new NFT named {\em Noun} (Figure \ref{nounsfig}) every once a day.
ii) Noun is automatically listed to the daily auction, which every peer can bid with Ether.   
iii) Noun holders can use pooled Ether to make Nouns more widespread (e.g., create T-shirts, pay to developers).\footnote{For the first five years, every tenth Noun auction is skipped, and the Nouns are automatically sent to the core developers. This mechanism serves as an incentive for the core developers, in lieu of pre-mined tokens.}
Since there is no centralized marketing manager, Nouns DAO needs a consensus-building among Noun holders on how to use the pooled Ether. 

For consensus-building, Nouns DAO adopts the token-voting mentioned above (\ref{strategic}).
Peers can vote for three choices ({\em for}, {\em against}, {\em abstain}) of the proposal with Nouns, and this action carries no penalty or reward (Figure \ref{nounsvote}).
Strategic behavior, spamming and Sybil attacks are prevented by this token-voting (albeit the risk of Keynesian beauty contest), but free-riding remains due to the lack of rewards. 
Consensus-building is currently driven by the enthusiasm of Nouns holders.\footnote{Nouns DAO employs a proxy voting system to mitigate low voter turnout; however, there are no rewards allocated for proxy voters.}
For token value, marginal cost is equivalent to the winning bid (Ether) in the daily auction, which is determined by the market.
As mentioned in \ref{tokendemand}, marginal utility exists as both an art and a governance token.
There is no stabilizer for the value of Nouns.

\begin{figure}[h] 
    \begin{minipage}[!t]{0.26\hsize}
    \centering
    \includegraphics[width=1.0\hsize]{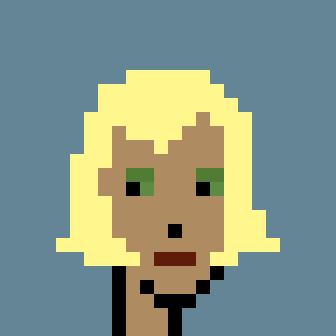} \\
    \vspace{1em}
    \end{minipage}
    \begin{minipage}[!t]{0.7\hsize}
    \centering
    \includegraphics[width=1.0\hsize]{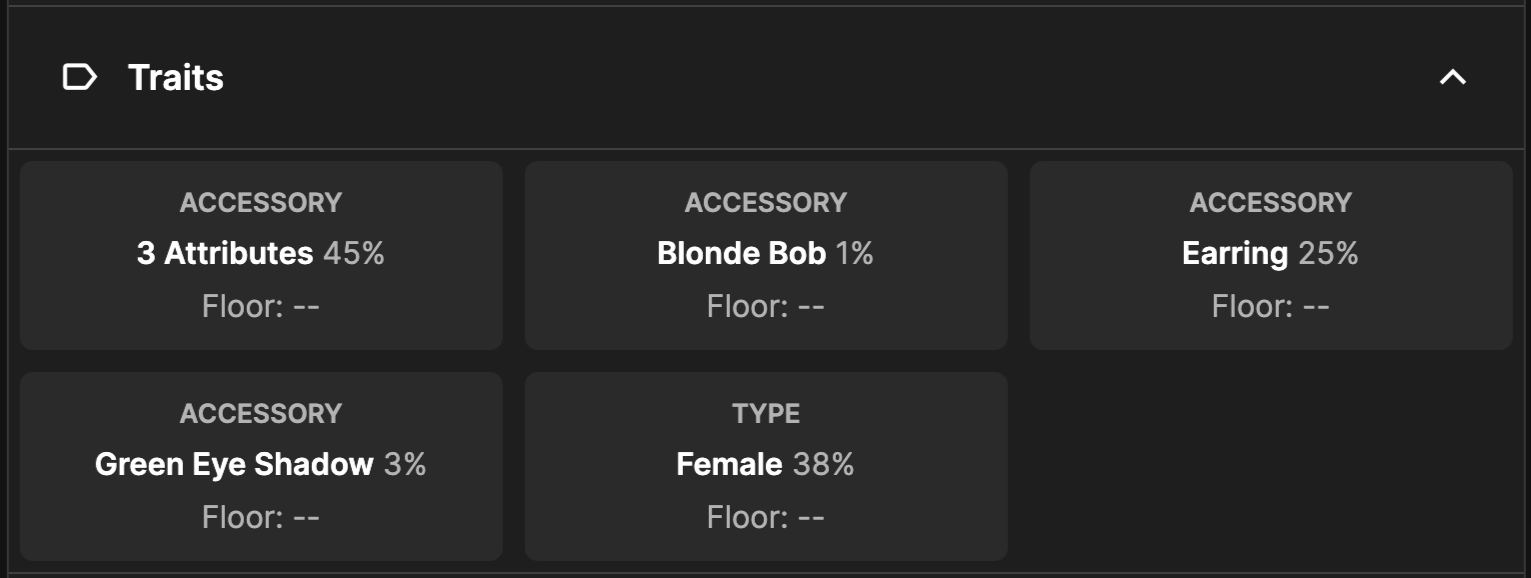} \\
    \vspace{1em}
    \end{minipage}
    \caption{Stochastic Rarity of NFT collections ({\em CryproPunks})}
    \captionsetup{font=footnotesize}
    \caption*{{\em Source}: OpenSea.io (\href{https://opensea.io/assets/ethereum/0xb47e3cd837ddf8e4c57f05d70ab865de6e193bbb/0}{https://opensea.io/assets/ethereum/0xb47e3cd837ddf8e\\4c57f05d70ab865de6e193bbb/0}, accessed November 8, 2023).}
    \captionsetup{font=normal}
    \label{rarity}                                      
\end{figure}

Nouns DAO, especially the idea of leveraging NFTs and daily auctions, is well-designed but does not address free-riding and lacks a stabilizer.
The author has two opinions on these issues.
First, adjusting the Nouns' rarity could solve free-riding.
Most NFT collections automatically generate groups of NFTs through programs that stochastically determine the traits of each part (Figure \ref{rarity}); 
NFTs with rare traits tend to have higher market prices \cite{mekacher2022rarity}.
Nouns are generated in the same way, but their stochastic elements are all uniformly distributed (i.e., there is no rarity in Nouns).\footnote{Nouns DAO generates pseudo-random numbers using the hash number of the previous block and the Nouns ID. More details can be found in the smart contract code at Etherscan. \href{https://etherscan.io/address/0xCC8a0FB5ab3C7132c1b2A0109142Fb112c4Ce515\#code}{https://etherscan.io/address/\\0xCC8a0FB5ab3C7132c1b2A0109142Fb112c4Ce515\#code}}
Here, the more a Noun is used in voting, the less likely its traits appear in the series of future Nouns.
This function would encourage Noun holders to participate in consensus-building.
Second, adjusting the auction frequency could be a stabilizer.
Nouns are currently generated once a day, but adjusting this frequency according to the previous winning bid (i.e., more frequent with higher bids, less frequent with lower bids) would help stabilize the marginal cost of Nouns.\footnote{This design is inspired by the difficulty adjustment. It would also be worth considering to reduce the auction frequency over time, similar to Bitcoin's block-reward halving, to increase the marginal cost.}
These NFT-specific extensions would make Nouns DAO more robust.\footnote{Recently, a fork occurred in Nouns DAO \cite{coindeskNounsDAOBarrels}, which implies the difficulty of designing decentralized autonomous consensus-building.}

\subsection{Terra} \label{caseterra}
Terra \cite{kereiakes2019terra} is a protocol for algorithmic stablecoins, where peers can build consensus on pegging and transaction records.
Terra consists of two tokens: TerraUSD (UST) and LUNA;\footnote{Terra issued stablecoins pegged to multiple currencies, although only USD is mentioned here for convenience.} the former is a stablecoin, and the latter is a coinbase for consensus-building on UST's transaction records.
LUNA works to stabilize UST prices by guaranteeing the exchange of one UST for one USD worth of LUNA.
If the market price of one UST falls below 1 USD, arbitrageurs can obtain profit by exchanging one UST for LUNA, which stabilizes the UST price because Terra burns received UST and mints LUNA (and vice versa).\footnote{In cases where LUNA is exchanged for UST, Terra only burns a portion of the received LUNA.}
That is, the UST price is intended to be stable indirectly through the supply adjustment of LUNA.\footnote{It is presumed that these tokens are named after the celestial phenomenon of the moon orbiting the Earth, symbolizing a balance similar to that of gravitational forces.}  

For consensus-building, Terra, as a protocol, adopts proof-of-stake (LUNA), Nakamoto consensus, and coinbase (LUNA).
The combination of proof-of-stake and Nakamoto consensus addresses strategic behavior and free-riding.
Spamming and Sybil attacks are prevented by the transaction fee (UST) and proof-of-stake.
For token value, LUNA has no marginal cost because it is minted automatically through exchange with UST and proof-of-stake based consensus-building.\footnote{The marginal cost of a token minted with proof-of-stake is the opportunity cost incurred by rendering the staked token unusable. For example, staking Ether means losing the opportunity to use it for DApps. However, since LUNA's utility is the staking itself, there is no such opportunity cost, resulting in no marginal cost for the new mintage of LUNA.}  
Marginal demand seems to be ensured by staking, i.e., peers can obtain LUNA (as coinbase) and UST (as transaction fee) due to consensus-building.
In addition to the above-mentioned UST-LUNA exchange, Terra has several stabilizers to make UST an algorithmic stablecoin; these include adjusting the amount of coinbase, transaction fees, and the burn rate of UST after exchange.

Terra appears to be well-designed, referring to the Bitcoin protocol;
however, this protocol collapsed in 2022 due to the inability to maintain the peg between UST and USD.
Once the price of 1 UST fell below 1 USD, the demand for exchange into LUNA surged, and in response, new LUNA was minted, which lowered the price of LUNA, making it difficult to exchange for the equivalent of 1 USD, further lowering the price of UST, and so on in a death spiral.
Several studies \cite{uhlig2022luna, briola2023anatomy, liu2023anatomy} empirically analyzed this incident, but from a design perspective, the fundamental problem is the lack of marginal costs in minting LUNA despite its use for price stabilization.

\subsection{Uniswap} \label{caseuni}
Uniswap \cite{uniswapHomeUniswap} is also a DApp on Ethereum,\footnote{"Uniswap is now usable on several protocols, including Ethereum, {\em Polygon}, {\em Optimism}, {\em Arbitrum}, and {\em BNB Chain}.} where peers can exchange tokens without a centralized entity (DEX).
In 2020, Uniswap began minting and distributing UNI, a governance token, to several types of users including liquidity providers \cite{uniswapIntroducing}.\footnote{\url{https://etherscan.io/address/0x1f9840a85d5af5bf1d1762f925bdaddc4201f984}}
Apart from AMM (Figure \ref{amm}), we review the governance of Uniswap using UNI tokens.

\begin{figure}[!t] 
    \begin{minipage}[!t]{0.45\hsize}
    \centering
    \includegraphics[width=1.0\hsize]{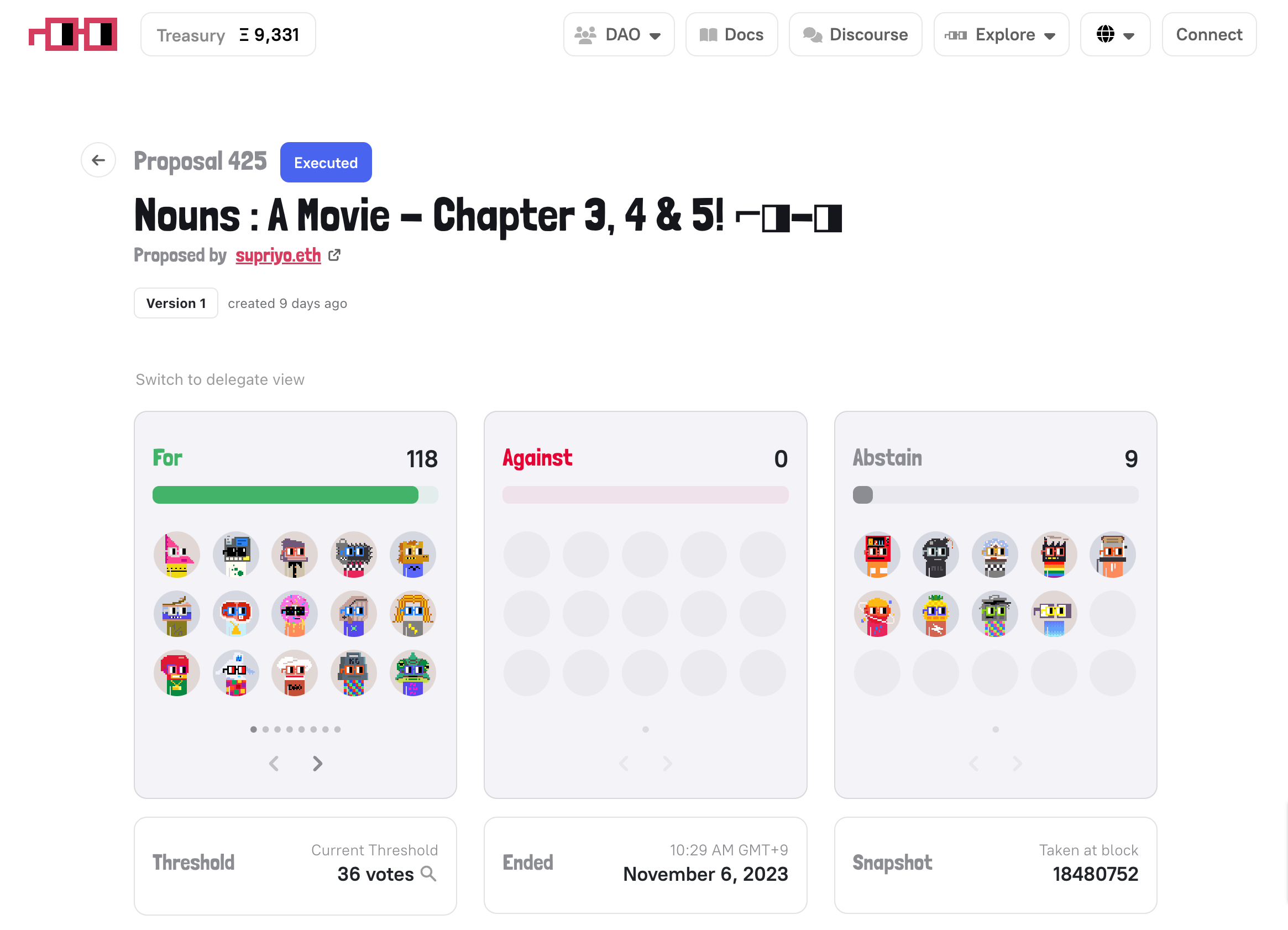} \\
    \captionsetup{labelfont=normalsize}
    \subcaption{Nouns DAO} \label{nounsvote}
    \vspace{1em}
    \end{minipage}
    \begin{minipage}[!t]{0.45\hsize}
    \centering
    \includegraphics[width=1.0\hsize]{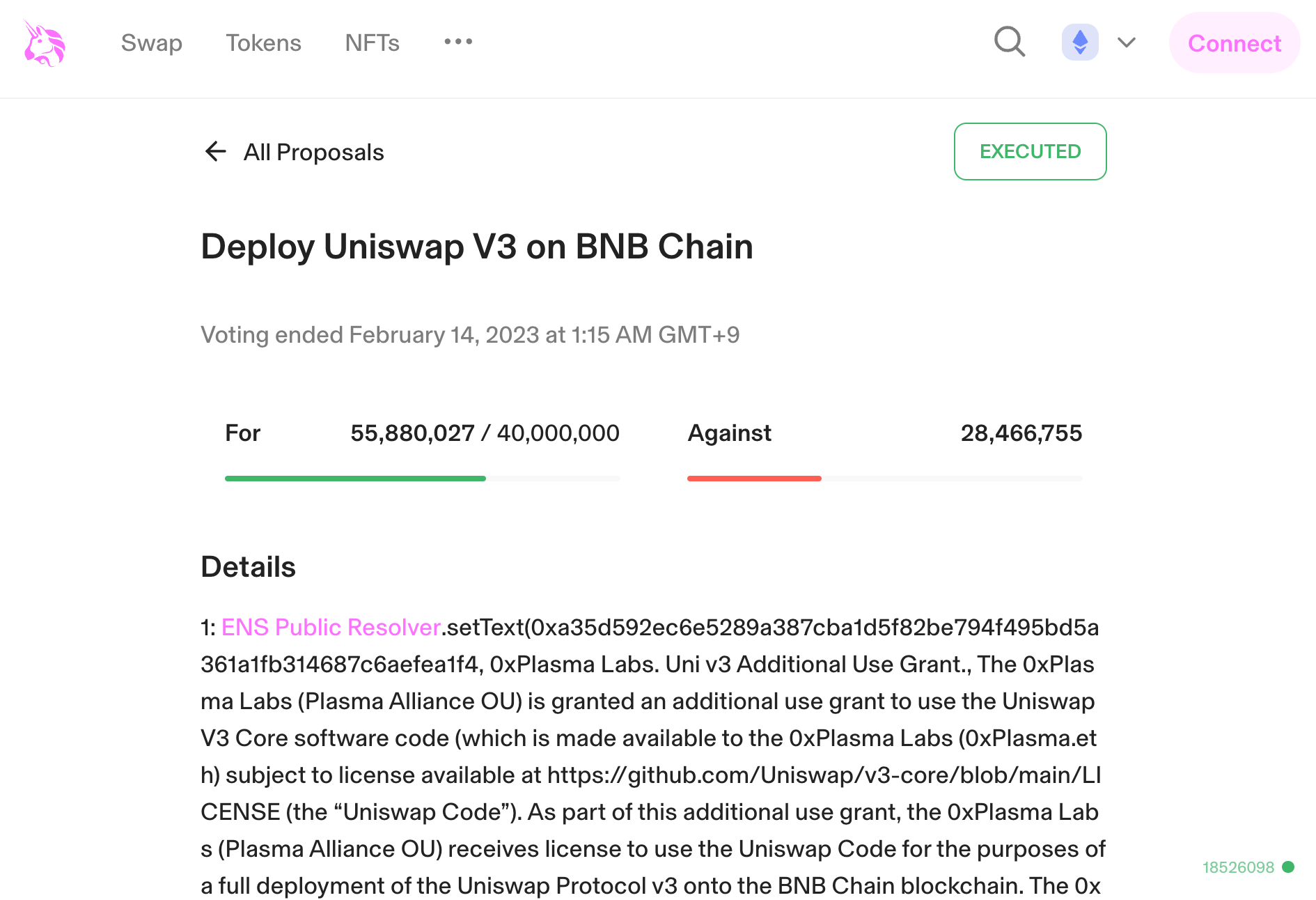} \\
    \captionsetup{labelfont=normalsize}
    \subcaption{Uniswap} \label{univote}
    \vspace{1em}
    \end{minipage}
    \caption{Voting with Governance Token} 
    \captionsetup{font=footnotesize}
    \caption*{{\em Sources}: (a) Nouns : A Movie - Chapter 3, 4 \& 5! (\url{https://nouns.wtf/vote/425}, accessed November 8, 2023), (b) Deploy Uniswap V3 on BNB Chain (\url{https://app.uniswap.org/vote/2/31}, accessed November 8, 2023).}
    \captionsetup{font=normal}
\end{figure}

Uniswap adopts token-voting with UNI for consensus-building, similar to that of Nouns DAO (Figure \ref{univote});
thus, strategic behavior, spamming and Sybil attacks are prevented but the risk of free-riding still exists.
Some studies \cite{barbereau2022defi, barbereau2023decentralised} indicate low voter turnout in the governance of Uniswap and other DEXs.
For token value, part of the marginal cost would be the opportunity cost of liquidity providing minus the reward through LP tokens (\ref{tokendemand}), given that UNI is periodically distributed to liquidity providers even after initial distribution.\footnote{Uniswap describes its periodic distribution of rewards as {\em liquidity mining}. Note that the opportunity cost of the liquidity mining includes the risk of impermanent loss. Impermanent loss refers to the potential loss a liquidity provider may incur due to fluctuations in the token ratio within the liquidity pool.}
Marginal utility is, for now, only the right to participate in governance.
There is no stabilizer for the value of UNI.

Uniswap is a representative DEX based on AMM, but its governance using UNI has free-riding challenges, a lack of stabilizer, and weak token value (i.e., weak marginal cost and marginal utility).
For the first two challenges, leveraging prior studies and products presented in \ref{free} and \ref{tokenstabilize} would be helpful.
For the last challenge, Uniswap must increases at least the marginal cost or utility of UNI.
A straightforward way would be to increase marginal utility by distributing a portion of the exchange fee to LP and UNI token holders (this extension, known as a {\em fee switch}, has long been controversial in Uniswap governance \cite{uniswapFeeSwitch, uniswapUniswapProposal}); however, the question arises whether UNI is necessary in the first place.
By analogy with stocks, we could design a simpler consensus-building if LP tokens were governance tokens.
Frankly, the raison d'etre of UNI is unclear.\footnote{In reality, the mintage of UNI tokens was a response to a {\em vampire attack}. This type of attack involves copying an open-source project and then trying to divert its resources by offering higher incentives. In 2020, Sushiswap \cite{sushiSushix1F363}, a clone of Uniswap, tried to attract a significant portion of Uniswap's liquidity by distributing its governance token, SUSHI. As a countermeasure, Uniswap was compelled to mint its own governance tokens, despite the unclear raison d'etre. See Fan et al. (2023) \cite{fan2023towards} for more details on this incident.}

\section{Conclusion} \label{conclusion}
This paper surveyed products and studies behind cryptoeconomics and tokenomics from an economic perspective, which aims to bridge the economic and blockchain contexts.

Our survey first organized the history of each term in chronological order, suggesting that cryptoeconomics and tokenomics can be novel when integrated (Section \ref{history}).
We then surveyed blockchain-related products and prior studies on designing consensus-building for decentralization (Section \ref{consensus}) and designing token value for autonomy (Section \ref{token}), respectively.
The former is a category related to cryptoeconomics, and the latter is to tokenomics.
Finally, from the integration viewpoint, we evaluated five products as a case study (Section \ref{case}).

These attempts illustrated the importance and difficulty of integration. 
Decentralized autonomous consensus-building requires at least simultaneous consideration of strategic behavior, spamming, Sybil attacks, free-riding, marginal cost, marginal utility, and stabilizers.
This task is complex and challenging for both research and implementation.
This survey aims to alleviate this difficulty as a first step in bridging the contexts of economics and blockchain.

Finally, two problems are worth mentioning for future research.

\subsection{How to Control External Incentives} \label{outside}

External incentives can also influence consensus-building.
For example, the Bitcoin protocol can be attacked even by peers who earn Bitcoin (coinbase) because there remains an incentive to lower the Bitcoin price as long as they can short-sell on external exchanges ({\em Goldfinger attack} \cite{kroll2013economics}). 
The application layer has recently influenced Ethereum (consensus-building at the protocol layer), as DEXs have created new revenue opportunities through reordering transactions within blocks ({\em maximal extractable value};\footnote{The term was originally {\em miner extractable value}, but this terminology has become outdated since the adoption of proof-of-stake in Ethereum led to the disappearance of miners.} MEV \cite{ethereumMaximalExtractable}).
Furthermore, governance issues are essential for protocols.
The Bitcoin protocol and Ethereum have experienced multiple blockchain splits (i.e., intended hard forks) due to the failure of consensus-building on the protocol design itself \cite{badari2021overview}.

\subsection{How to Alleviate Rationality} \label{irrational}

This survey consistently assumed that peers are rational, i.e., they act to maximize the expect amount of reward tokens (of sufficient value);
however, consensus-building would be more robust and practical if it were not always considered rational.
Prior studies on {\em behavioral economics} may be helpful to this perspective.
Behavioral economics is already being applied to blockchain-related discussions, from market analysis \cite{al2020cryptocurrency} to DApps design \cite{toyoda2023web3}, and synergies with cryptoeconomics have also been noted \cite{mediumBehavioralCryptoEconomics}.

\vspace{0.4\baselineskip}

\noindent
Addressing these open problems will be necessary to design decentralized autonomous consensus-building.

\section*{Acknowledgment}

The author would like to express his gratitude to Mohammed bin Salman Center for Future Science and Technology for Saudi-Japan Vision 2030 (MbSC2030) for providing financial support.

\bibliography{refs} 
\bibliographystyle{IEEEtran} 

\end{document}